\documentstyle[prl,multicol,aps,epsf]{revtex}

\begin{document}
\title{Spin Singlet Mott States and Evidence For Spin Singlet Quantum 
Condensates 
of Spin-One Bosons in Lattices}
\author{Fei Zhou$^{\dagger,\dagger\dagger}$ and Michiel Snoek$^{\dagger}$}
\address{ITP, Utrecht University, Leuvenlaan 4,
3584 CE Utrecht, The Netherlands$^{\dagger}$\\
Department of Physics and Astronomy,
University of British Columbia,\\
6224 Agriculture Road, Vancouver, B.C. V6T 
1Z1,Canada$^{\dagger\dagger}$\footnote{ Permanent address}}
\date{\today}
\maketitle

\begin{abstract}

We have investigated spin singlet Mott states
of spin-one 
bosons with antiferromagnetic interactions.
These spin singlet states do not break rotational symmetry and
exhibit remarkably different
macroscopic properties compared with nematic Mott states
of spin-one bosons. 
We demonstrate that the dynamics 
of spin singlet Mott states
is fully characterized by even- or odd-class
quantum dimer models.
The difference between spin singlet Mott states for even and odd numbers 
of atoms per 
site can be attributed to a selection rule in the low energy sectors of 
on-site Hilbert spaces; alternatively, it can also be attributed to 
an effect of Berry's phases on bosonic Mott states.

We also discuss evidence for
spin singlet quantum condensate of spin-one atoms.
Our main finding is that
in a projected spin singlet Hilbert space, the low energy physics
of spin-one bosons is equivalent to that of a Bose-Hubbard model for
{\em spinless} bosons interacting via Ising gauge fields.
The other major finding is spin-charge 
separation in some one-dimensional Mott states.
We propose charge-e 
spin singlet superfluid 
for an odd 
number of
atoms per lattice site 
and charge-2e spin singlet superfluid for an even number of atoms 
per lattice site in one-dimensional lattices.
All discussions in this article are limited to integer numbers of bosons 
per site.
\\
\\
{PACS number: 03.75.Mn,05.30.Jp,
75.10.Jm}
\end{abstract}

\begin{multicols}{2}

\narrowtext

\section{Introduction}
Spin correlated condensates of spin-one atoms have attracted 
considerable interest since 1998\cite{Stamper-Kurn98,Stenger98}.
Two-body scattering between atoms leads to either antiferromagnetic 
(polar)
or ferromagnetic condensates\cite{Ho98,Ohmi98,Law98}.
For spin-one bosons with antiferromagnetic interactions,
the condensates break both the rotational symmetry and
the $U(1)$ symmetry associated with the phases of the condensates. There are
two branches of spin-wave excitations and
one gapless phason mode. 
For a condensate of a finite size, the exact ground state
however is a spin singlet (assuming the total number of atoms is
even ) which on the other hand is extremely sensitive to external 
perturbations\cite{Law98,Ho00}.
In rotating traps,
two-body scatterings furthermore result in spin correlated fractional
quantum Hall states\cite{Ho,Schoutens}.

In lattices where the hopping is small, bosonic particles can be 
localized
because of repulsive interactions and ground states in 
this
limit are Mott insulators\cite{Fisher89}.
Mott-insulating states of cold atoms have been recently observed in
optical lattices\cite{bloch01}.
Spin correlated Mott-insulating states in high dimensional optical 
lattices have also been investigated theoretically and reported in
\cite{Demler01}.
Both nematic Mott insulators which break the rotational symmetry and 
spin singlet Mott insulating states 
were found for high dimensional lattices in certain parameter regime.
Other spin correlated states which break both translational and
rotational symmetries were 
proposed in\cite{Yip03}.
Effects of spin correlations on Mott insulator-superfluid 
transitions were suggested and remain to be
fully understood\cite{Tsuchiya02}.
Very recently,
detailed analysises of microscopic wave functions and phase transitions 
between
spin singlet Mott insulators and nematic ones have been carried 
out\cite{Demler03,Snoek03}.
In the context of 
antiferromagnets, the issue of
spin nematic states 
was raised and addressed 
previously\cite{Andreev84,Chubukov90,Larkin90}.

To distinguish different correlated states of spin-one bosons in a 
lattice, it is
useful to introduce the following order parameters 
in terms of creation and annihilation operators $\psi^\dagger_\alpha$, 
$\psi_\alpha$ respectively ($\alpha=x,y,z$);

\begin{eqnarray}
&& {\cal O}^1_\alpha=\langle \psi_{k\alpha}\rangle , \nonumber \\
&& {\cal O}^2_{\alpha\beta}=\langle \psi^\dagger_{k\alpha}\psi_{k\beta}\rangle 
-\frac{1}{3}\delta_{\alpha\beta}\langle \psi^\dagger_{k\gamma}\psi_{k\gamma}\rangle .
\label{op}
\end{eqnarray}
In Eq.\ref{op},
we have employed the following creation
operators

\begin{eqnarray}
&& {\bf \psi}^\dagger_x=\frac{1}{\sqrt{2}}\big(\psi^\dagger_{-1} 
-\psi^\dagger_{1}\big),\nonumber \\
&& {\bf \psi}^\dagger_y=\frac{i}{\sqrt{2}}\big( \psi^\dagger_1 
+\psi^\dagger_{-1}\big),\nonumber \\
&& {\bf \psi}^\dagger_z=\psi^\dagger_0.
\end{eqnarray}
where ${\bf \psi}^{\dagger}_{m_F}$, $m_F=\pm 1,0$, are creation operators of spin- 
one bosons with $S_z=m_F$.

In terms of these order parameters,
polar condensates (pBEC), nematic Mott insulators(NMI) and
spin singlet Mott insulating states (SSMI) have the following 
properties

\begin{eqnarray}
&& {    \text{pBEC}:     } {\cal O}^1_\alpha \neq 0,
{\cal O}^2_{\alpha\beta}\neq 0;
\nonumber
\\
&& {  \text{NMI}:    } {\cal O}^1_\alpha=0,
{\cal O}^2_{\alpha\beta}\neq 0;
\nonumber\\
&& { \text{SSMI}: }  {\cal O}^1_\alpha=0,
{\cal O}^2_{\alpha\beta}=0.
\end{eqnarray}

Finally,
for all states of cold atoms with antiferromagnetic interactions
which interest us in this paper,
the expectation value of spin operators is zero 

\begin{equation}
\langle {\bf S}\rangle =0, 
{\bf 
S}^\alpha= -i\epsilon^{\alpha\beta\gamma}\psi^\dagger_\beta \psi_\gamma;
\alpha,\beta,\gamma=x,y,x.
\end{equation}

Spin correlations in Mott-insulating states also depend on the 
even-odd 
parity of numbers of atoms per site.
While spin correlated Mott states for even numbers of atoms per site
have been studied and understood, states for odd numbers of atoms per 
site remain to be fully understood. In one-dimensional optical 
lattices, it was argued that spin singlet Mott states for 
an odd number of atoms per site should be dimerized-valence-bond 
crystal (DVBC) states\cite{Zhou3,Demler03}. These DVBC states are 
characterized by 
the following nontrivial spin correlations:

\begin{eqnarray}
&& \langle {\bf S}_i\cdot{\bf S}_{i+1}{\bf S}_{i+2k}\cdot{\bf S}_{i+2k+1}\rangle 
\neq 0
\end{eqnarray}
when $k$ approaches $\infty$ (for either all even $i$ or all odd $i$ 
only).
Later, we present more evidence for the existence of
DVBC Mott states;
certain aspects of long wave length behaviors of
DVBCs found for spin-one bosons are similar to those of DVBC states in
$S=1$ spin 
chains\cite{Affleck87,Affleck89,Arovas88,Parkinson88,Barber89,Chubukov91}.

Valence-bond crystal states and more generally spin-Peirls states have 
been proposed in various models for strongly correlated electrons.
Possible spin-Peirls states in 2D antiferromagnets
(disordered) due to Berry's phases carried by  
hedgehogs were pointed out in \cite{Haldane88}.
In {\em SU(N)}-antiferromagnet models or their effective models, valence-
bond crystal phases were discovered in various 
limits\cite{Read89,Read91,Jalabert91}. 
Most recently, valence-bond crystals in strongly-correlated systems due 
to Berry's phases in Ising gauge fields
were addressed\cite{Senthil99,Senthil00}.
In quantum dimer models,
crystal phases have been anticipated in\cite{Rokhsar88}, and were reviewed 
in a recent paper\cite{Moessner01}.

Condensates for spin-one bosons with antiferromagnetic interactions
in lattices also have fascinating properties;
most of arguments in the weakly interacting case can be carried out 
parallel to those developed
for the continuous limit. Particularly,
ground state wave functions 
should possess Ising gauge symmetries; and condensates 
support 
interesting half-vortex-type topological excitations in additional to
usual integer vortices\cite{Zhou01}. Furthermore,
the local spin dynamics is described by 
a nonlinear sigma model;
for an individual condensate in a zero-mode
approximation, a constrained $O(3)$-quantum-rotor model can be introduced
to characterize the spin dynamics.
Due to spin-phase separation there are
a variety of novel spin correlated "condensates";
particularly, 
we will argue that
there should be spin singlet quantum condensates (SSQCs),
in addition to usual polar condensates.

Early indication for spin singlet quantum condensates, or rotationally
invariant condensates
came from the analysis of renormalization-group 
equations.
The phenomenon of
spin-phase separation 
implies that
orders be established in
either phase, or spin sector but not necessary in both. 
Particularly, there exist rotationally invariant superfluid 
states\cite{Zhou01}.
In fact,
the failure of order parameters 
(when treated as quantum operators)
to commute with the Hamiltonian 
eventually leads to these fascinating 
many-body condensates of spin-one bosons.
From this point of view,
a polar condensate is a "classical" one, of which order parameter 
operators
can be approximated as classical variables.

The second relevant observation on spin singlet condensates 
was made in \cite{Demler01}. There, the authors consider
a limit where two-body repulsive interactions in ferromagnetic 
channels are much stronger than in singlet channels.
In this limit, atoms form singlet pairs and condense;
the resultant state is a condensate of spin singlet pairs.
The solution in this limit is consistent with 
general arguments based on renormalization-group equations;
furthermore, it indicates a microscopic realization of rotationally 
invariant condensates suggested by renormalization-group equations.
In \cite{Liu02}, the authors further analyzed a few other  
spin singlet states, some of which break time reserval symmetry and
have fascinating edge properties.

The most recent indication of SSQCs is the phenomenon of 
fractionalization in some Mott states.
Consider each cold atom as a particle with spin $S=1$ and "charge" $Q=e$.
The spin-"charge" separation 
in Mott insulating states for spin-one bosons with antiferromagnetic
interactions 
implies fractionalization of cold atoms in optical
lattices;  
it also offers some insight into microscopic wavefunctions
of fractionalized condensates.
The ground state of interacting spin-one bosons
in the strong coupling limit can be a DVBC; and this state 
supports either spinful but chargeless quasi-particles or
charged but spinless quasi-particles in one-dimensional limit\cite{Zhou3}.
The low-energy sector of the Hilbert space can thus be divided into two
subsectors of spin and charge. All the 
low lying excitations live in one of these two sectors. 
Therefore, each atom injected into a lattice naturally fractionalizes
into spinless {\em charge-e} ($Q=1, S=0$) excitations and chargeless but 
spinful ($Q=0, S=1$) 
excitations.
Theoretically, the novel SSQCs  
appear naturally as a result of condensation of spinless charge-e bosons.
In this article, we substantialize those proposals 
made in \cite{Zhou3}; particularly we demonstrate the microscopic
structure of possible charge-e spin singlet condensates \cite{Lorentz02}.

The fractionalization of cold atoms to be discussed
in this article is an analogy of spin-charge separation proposed in
strongly interacting electron systems.
For strongly-correlated electrons in cuprates,
the concept of spin-charge separation was emphasised  
in \cite{Anderson87};
the phenomenon of spin-charge separation in doped anti-ferromagnets
was studied afterwards in terms of effective compact gauge 
fields \cite{Anderson88,Anderson87a}.
Many attempts have since then been made to 
understand
the fractionalization of electrons in 
cuprates
\cite{Read89,Rokhsar88,Affleck88,Ioffe89,Wen89,Wen89a,Lee89,Fradkin90,Read91a}.
In some subsequential works, 
fractionalization in quantum spin liquids  
was investigated in the context of effective Ising gauge theories
\cite{Read91,Jalabert91,Wen91}. Most 
recent efforts on this subject can be found in 
\cite{Balents98,Senthil99,Senthil00,Moessner00,Sachdev00,Moessner01,Demler02};
especially, superconducting states of "chargons" (charge-e objects) have 
been proposed in \cite{Senthil99,Demler02}.
Some general discussions on spin-charge separation can be found 
in a remarkable book \cite{Anderson97}.

It is worth pointing out that
from the point of view of fractionalized atoms,
in some Mott insulating states only a fraction 
of each spin-one boson
appears to be "localized"; these states display rather interesting 
internal coherence as a result of "partial condensation". For instance,
nematic Mott insulating states can be effectively considered as 
condensates of fractionalized 
spinful but chargeless particles or spin-one {\em spinons}.
In appendix C, we examine spin correlated Mott states from this 
point of view.
Our discussions will be limited to cases of integer numbers of bosons per 
site; condensates for non-integer numbers of bosons per site
will be addressed in a future paper.

In this article, we develop a general approach for the studies of
spin singlet states. First, we demonstrate that all
SSMIs can be fully characterized by even- and odd-class quantum dimer 
models.
In the second part of the article, we pursue the idea of 
SSQCs of spin-one bosons in lattices; particularly, we provide 
various evidence for these unconventional spin correlated condensates. 
In SSQCs, 
condensation takes place only in the charge sector of the
Hilbert space. SSQCs therefore exhibit distinct long-range order.  It is
possible to distinguish them from pBECs in experiments because of remarkably
distinct macroscopic properties of SSQCs.

In section II, we 
summarize results on spin-
correlated Mott states of spin-one bosons in lattices
and
present microscopic wavefunctions
for some of the states.
In section III, 
we investigate spin singlet Mott states for odd 
numbers of
atoms per site in one-dimensional lattices
in the extremely small hopping limit.
In section IV, we discuss excitations in spin singlet Mott states;
we demonstrate the fractionalization of spin-one atoms in DVBCs
in one-dimensional lattices.
In section V, we present a qualitative picture of SSQCs using a projected 
spin singlet Hilbert space; we argue that spin correlations
strongly suppress one-particle hopping in an SSMI for an even
number of bosons per site.
In section VI, we introduce a fractionalized representation for
interacting spin-one bosons in 
a lattice. 
In section VII, we briefly present results on SSMIs and SSQCs in high 
dimensional 
lattices.
In section VIII, we revisit low-dimensional SSMIs using an approach based 
on quantum dimer models;
particularly, we generalize the results on one-dimensional lattices in 
section III to the entire 
Mott phase (one-dimensional).
In section IX, we 
discuss SSQCs in one-dimensional lattices. 
In section X, we address the issue of the possible role of Berry's phases 
in one-dimensional lattices. 

It was demonstrated 
and emphasized in a few occasions that 
the problem of interacting spin-one bosons can be mapped into
a constrained quantum rotor model (CQR) \cite{Zhou01,Demler01}. 
Technically, this allows one to acquire certain valuable intuition 
about many strongly-correlated states of spin-one bosons using the
knowledge about the renormalization-group equations and topological field 
theories. The mapping is therefore a powerful device to study 
properties of spin-one bosons in optical lattices.
The resultant CQR model will be our starting point for the studies 
of spin-one bosons in lattices.

\section{Spin Correlated Mott States:\\ A short review}

For spin-one bosons with antiferromagnetic interactions  
in (optical) lattices, the effective Hamiltonian 
can be conveniently expressed as
\begin{eqnarray}
&& {\cal H}_{lat.}=
-\tilde{t} \sum_{\langle kl\rangle }({\bf \psi}^\dagger_{k\alpha} {\bf \psi}_{l\alpha} + 
h.c.)+
\sum_{k} \frac{{\bf S}_k^2}{2I} + \frac{{\hat{\rho}}_k^2}{2C}
-\hat{\rho}_k\mu
\label{hami0}  
\end{eqnarray}
where the sum $\langle kl\rangle $ is carried out over neighboring sites $k$ and $l$; in this 
section, we are only interested in high dimensional bipartite lattices.

In Eq.\ref{hami0},
$E_s=1/2I$ and $E_c=1/2C$ are two energy gaps in the excitation spectra 
of an individual 
condensate at each site. $\tilde{t}$ is the hopping integral;
$\mu$ is the chemical potential. $N_0(\gg 1)$  will be the 
number of bosons per site.
The dynamics is defined by the following commutation relations,

\begin{eqnarray}
&& [{\bf S}^{\alpha}_k, {\bf \psi}_{l\beta}]=
i\epsilon^{\alpha\beta\gamma}\delta_{kl} {\bf \psi}_{k\gamma},
[\hat{\rho}_k, {\bf \psi}^\dagger_{l\alpha}]=\delta_{kl}\delta_{\alpha\beta}
{\bf \psi}^\dagger_{l\beta}
\label{Algebra}
\end{eqnarray}
where 

\begin{eqnarray}
&&{\bf S}_k^\alpha=-i\epsilon^{\alpha\beta\gamma}{\bf 
\psi}_{k\beta}^\dagger{\bf 
\psi}_{k\gamma}, {\rho}_k={\bf \psi}^\dagger_{k\alpha}{\bf \psi}_{k\alpha} 
\end{eqnarray}
are the spin and 
"charge" operators for spin-one 
bosons, $\alpha (\beta,\gamma)=x, y,z$. And 
\begin{eqnarray}
&&[{\bf S}^\alpha_k, {\bf S}^\beta_{k'}]=i\epsilon^{\alpha\beta\gamma}
\delta_{k,k'}{\bf S}^\gamma_k,
[{\bf S}^{\alpha}_k, \hat{\rho}_{k'}]=0\nonumber \\
&& [{\bf S}^{\alpha}_k, \psi^\dagger_{k'\beta}\psi^\dagger_{k'\beta}]
=[{\bf S}^{\alpha}_k, \psi_{k'\beta}\psi_{k'\beta}]=0
\end{eqnarray}
In addition,

\begin{equation}
{\bf S}_k^2=\hat{\rho}_k(\hat{\rho}_k+1)-
\psi^\dagger_{k\alpha}\psi^\dagger_{k\alpha}\psi_{k\beta}\psi_{k\beta}
\end{equation}
where $\psi_\alpha^\dagger\psi_\alpha^\dagger$ is a singlet pair creation 
operator.
Following this identity, ${\bf S}_k^2=S_k(S_k+1)$ and
$S_k=0,2,4,...,2n$, if $\rho_k=2n$; otherwise
$S_k=1,3,5,...,2n+1$ if $\rho_k=2n+1$.
For spin-one bosons with antiferromagnetic interactions,
the ground state for an isolated site with 
$\rho_k=2n$ atoms is a spin singlet 
and for $\rho_k=2n+1$ is a spin triplet\cite{Law98}.

Therefore
at low energies,
the Hilbert space of the Hamiltonian is subject to a constraint that 
the sum of the number of particles  $\rho_k$ and the total spin $S_k$,
has to be an even number so that 
the many-boson wave function is symmetric under interchange of two
spin-one particles. Say it differently, the 
Hilbert space displays
the following selection rule

\begin{equation}
(-1)^{S_k+\rho_k}=1 (S_k \leq \rho_k).
\label{selec}
\end{equation}
This {\bf selection rule} plays an extremely important role in both 
symmetry 
breaking 
states and spin singlet Mott insulating 
states\cite{Demler01,Demler03,Snoek03,Zhou3,Zhou01}.

In optical lattices, the hopping $\tilde{t}$ 
can be varied continuously as a 
function of laser intensity; $\tilde{t}$ decreases as the laser intensity 
increases. 
$E_{s,c}$ on the other hand depend on the scattering lengths $a_F$, $F=0,2$

\begin{equation}
E_{s}=\frac{4\pi \rho_0 (a_2-a_0)}{3M N_0},
E_c=\frac{4\pi\rho_0(2a_2+a_0)}{3M N_0 }.
\end{equation}
We will focus on the case when $E_s$ is much less than $E_c$.
As laser intensities are varied, we then have at least three 
different situations;

\begin{eqnarray}
&& \text{i)} t(V_{opt}) >> E_c >> E_s; \nonumber \\
&& \text{ii)} E_c >> t(V_{opt}) >> \sqrt{E_c E_s}; \nonumber \\
&& \text{iii)} E_c >> \sqrt{E_c E_s} >> t(V_{opt}).
\end{eqnarray}
In Eq.13, $t=N_0\tilde{t}$ is the effective hopping matrix element
including a standard bosonic factor $N_0$.
The wave functions of spin correlated states we are interested in, can be 
conveniently expressed in a coherent state representation

\begin{equation}
|{\bf n},\chi\rangle =\frac{1}{\sqrt{2\delta N}} \sum_{m=N_0-\delta N}^{N_0+\delta N}
\exp(-im\chi) \frac{\big(\chi^\dagger_\alpha {\bf 
n}_\alpha\big)^m}{\sqrt{2(m-1)!}} |0\rangle ,
\end{equation}
with $\delta N \ll N_0$.

So at each site, one introduces 
two "collective" coordinates ${\bf n}_k$ and $e^{i\chi_k}$, 
two unit vectors on a two-sphere $S^2$ and a unit circle $S^1$ 
which characterize the orientation of 
$O(3)$ and $O(2)$ quantum rotors respectively.
Following Eqs.6-11, the Hamiltonian ${\cal H}_{\text{lat.}}$ ($N_0,\delta N 
\gg 
1$) in this representation can be expressed as ${\cal H}_{\text{CQR}}$ of a
constrained quantum rotor (CQR) model

\begin{eqnarray}
&& {\cal H}_{\text{CQR}}=-t\sum_{\langle kl\rangle } {\bf n}_k\cdot {\bf n}_l 
\cos(\chi_k-\chi_l)
\nonumber\\
&& +\sum_k E_s {{\bf S}_k^2} +{E_c}{\rho_k^2}-\rho_k \mu.
\end{eqnarray}
And 

\begin{eqnarray}
&& [{\bf S}^\alpha_{k'},
{\bf n}^\beta_k] =-i\epsilon^{\alpha\beta\gamma}
\delta_{k,k'}{\bf n}^\gamma_k, {\bf 
S}^\alpha_k=i\epsilon^{\alpha\beta\gamma}{\bf n}^\beta_k 
\frac{\partial}{\partial {\bf n}_k^\gamma};
\nonumber \\
&& [\rho_{k'},\chi_k]=i\hbar\delta_{kk'}, 
\rho_k=i\hbar\frac{\partial}{\partial \chi_k}.
\end{eqnarray}

As the orientation of the $O(3)$-rotor at site $k$ is specified as ${\bf 
n}_k$,
the total spin at each site is defined as the angular momentum of this  
rotor, following Eq.15.
The phase $\chi_k$ further defines the orientation of an $O(2)$-rotor; the 
number operator corresponds to its angular momentum.
The sum of $S_k$ and $\rho_k$ again is subject to the constraint 
in Eq.11. 
The Hamiltonian ${\cal H}_{\text{CQR}}$ is locally invariant under a 
simultaneous inversion of ${\bf n}_k$ and $\exp(i\chi_k)$. Introducing 
operators:
\begin{eqnarray*}
\hat O_k^{\chi} \Psi(\ldots, \chi_k, \ldots) = \Psi(\ldots, \pi+\chi_k, 
\ldots) \\
\hat O_k^{\bf n} \Psi(\ldots, {\bf n}_k, \ldots)= \Psi(\ldots, -{\bf n}_k, 
\ldots )
\end{eqnarray*}
this invariance can be expressed as:
\begin{equation}
(\hat O_k^{\chi} \hat O_k^{\bf n})^{-1} 
{\cal H}_{\text{CQR}}
\hat O_k^{\chi} \hat O_k^{\bf n}
={\cal H}_{\text{CQR}}.
\end{equation}

In the limit i),
according to Eq.15, a usual polar condensate is 
established. 
A "mean field" polar condensate in this representation is

\begin{eqnarray}
&&\Psi_{\text{pBEC}}(\{ {\bf n}_k\}, \{\chi_k\}) 
 \propto \prod_{k} \delta(
 ({\bf n}_k \cdot 
{\bf n}_0 ) e^{i(\chi_k -\chi_0)}-1)
\label{pbec}
\end{eqnarray}
where all directors ${\bf n}_k$ point in the direction of ${\bf n}_0$ and
all $\chi_k$ are "locked" at the same value of $\chi_0$.
The wave function in Eq.\ref{pbec} explicitly exhibits the local Ising gauge 
symmetry emphasised in \cite{Zhou01}, i.e.

\begin{equation}
\hat O_k^{\bf n} \hat O_k^{\chi}
\Psi_{\text{pBEC}}=  \Psi_{\text{pBEC}}.
\end{equation}
Microscopically, 

\begin{eqnarray}
&& \Psi_{\text{pBEC}}\propto
\frac{(\chi^{\dagger}_{Q=0,\alpha}{\bf n}_{0\alpha}\exp(i\chi_0))^{N_0\times 
V_T}}{\sqrt{(N_0\times
V_T)!}} |0\rangle 
\end{eqnarray}
represents a condensate at a state with zero crystal momentum $Q=0$, along 
axis ${\bf n}_0$;
$\chi^{\dagger}_{Q\alpha}=\frac{1}{\sqrt{V_T}}\sum \exp(ikQ)\chi^{\dagger}_{k\alpha}$,
here $V_T$ is the number of sites in a lattice.

In the limits of ii) and iii), particles are localized at each site and
there are $N_0$ atoms per site. The wavefunction can be factorized into
\begin{equation}
\Psi(\{{\bf n}_k\},\{\chi_k \})
=\Psi_{N_0}(\{{\bf n}_k\})\otimes 
\prod_k \frac{1}{\sqrt{2\pi}}\exp(iN_0\chi_k)
\end{equation}

\begin{figure}
\begin{center}
\epsfbox{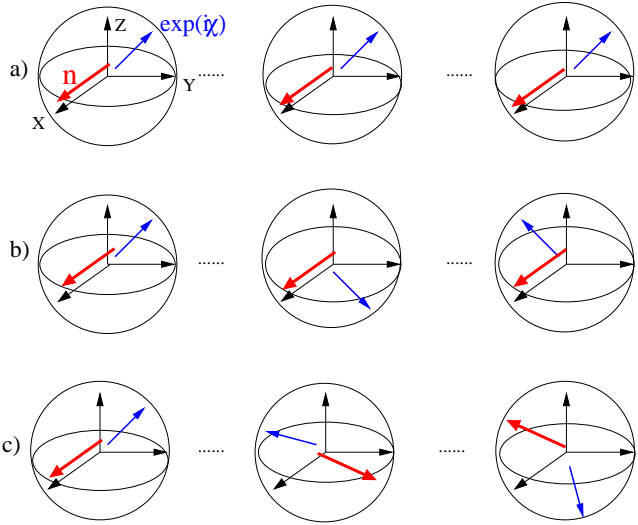}
\leavevmode
\end{center}
\caption{Ordering of two order parameters, unit vectors ${\bf n}$
(a light
arrow defined
on a two sphere $S^2$) and
$\exp(i\chi)$ (a dark arrow defined on a unit circle $S^1$,
drawn in the $YZ$-plane of the two sphere)
in different correlated states. a)pBEC, b)NMI and c) SSMI.
Two spheres and unit circles ($S^2$ and $S^1$) for a few different lattice
sites are shown here.}
\end{figure}

In terms of collective 
coordinates ${\bf n}_k$, the effective Hamiltonian for
Mott insulating states can be expressed as

\begin{eqnarray}
&& {\cal H}_{\text{MI}}=-
J_{ex}\sum_{\langle kl\rangle } Q_{\alpha\beta}({\bf n}_k)
Q_{\alpha\beta}({\bf n}_l)+E_s\sum_k {{\bf S}_k^2};
\nonumber \\
&& Q_{\alpha\beta}({\bf n})={\bf n}_\alpha{\bf n}_\beta-
\frac{1}{3}\delta_{\alpha\beta}.
\label{Hmi}
\end{eqnarray}
The exchange interaction $J_{ex}$ is of the order of ${t^2}{E_c}^{-1}$.
The Hilbert space observes the following 
symmetry under a local inversion $\hat{O}_k^{\bf n}$, 

\begin{equation}
\hat{O}^{\bf n}_k \Psi_{N_0}=(-1)^{N_0}\Psi_{N_0};
\end{equation}
Finally,
the Hamiltonian is invariant under the local inversion
\begin{equation}
(\hat{O}_k^{\bf n})^{-1}{\cal H}_{\text{MI}}
\hat{O}_k^{\bf n} ={\cal H}_{\text{MI}}.
\end{equation}

In the limit of ii),
one can identify the ground state as a nematic Mott insulator. 
In this intermediate regime, the director
${\bf n}$ is localized on a two-sphere $S^2$.
The wavefunctions for an even (e) and odd (o) number of atoms per site 
are

\begin{eqnarray}
&& \Psi^e_{\text{NMI}}(\{ {\bf n}_k\})\approx 
\prod_k \delta( ({\bf n}_k \cdot {\bf n}_0)^2-1) \nonumber \\
&& \Psi^o_{\text{NMI}}(\{ {\bf n}_k\})\approx
\prod_k \delta( ({\bf n}_k \cdot {\bf n}_0)^2 -1 ) 
({\bf n}_k\cdot {\bf n}_0).
\label{eo}
\end{eqnarray}
One notices that
\begin{eqnarray}
&& \hat{O}_k^{\bf n} \Psi^e_{\text{NMI}} =\Psi^e_{\text{NMI}},
\nonumber \\
&& \hat{O}_k^{\bf n} \Psi^o_{\text{NMI}}=-\Psi^o_{\text{NMI}}
\label{eo1}
\end{eqnarray}
so that the constraint in Eq.23 is satisfied.
Eq.\ref{eo} indicates that at each site all localized atoms condense 
in an identical spin-one state characterized by $\psi^\dagger_\alpha {\bf 
n}_{0\alpha}$; microscopically (see Fig. 2a))

\begin{equation}
\Psi^{e,o}_{\text{NMI}}\approx \prod_{k} \frac{(\chi^\dagger_{k\alpha}{\bf 
n}_{0\alpha})^{N_0}}
{\sqrt{N_0!}} |0\rangle 
\end{equation}
which corresponds to a maximally ordered state.

A direct calculation of order parameters defined in Eq.\ref{op}
in section I yields 

\begin{eqnarray}
&&{   \text{pBEC}:     } {\cal O}^1_\alpha=\sqrt{N_0}{\bf n}_\alpha 
\exp(i\chi_0); \nonumber 
\\
&&{   \text{NMI}:    } {\cal O}^1_\alpha=0,
{\cal O}^2_{\alpha\beta}=N_0
\big( {\bf n}_\alpha{\bf n}_\beta-\frac{1}{3}\delta_{\alpha\beta}\big).
\end{eqnarray}
Nematic Mott insulating states can be observed
at temperatures lower than the exchange energy $J_{ex}$.

Nematic states break the rotational symmetry. The internal 
space and its first homotopy group are

\begin{eqnarray*}
{\cal R}=\frac{S^2}{Z_2}, \pi_1({\cal R})=Z_2.
\end{eqnarray*}
These states support interesting $\pi$-spin disclinations
(some general discussions on the effects of Ising symmetry on atomic 
defects
can be found in an early work \cite{Zhou01}).

In the limit of iii), ${\bf n}$ and $\chi$ are delocalized on the
unit sphere $S^2$ and unit circle $S^1$ respectively; for an even number ($N_0=2n$) of particles per 
site, the wave function for a spin singlet Mott insulator (SSMI) can be written 
as 

\begin{equation}
\Psi^{e}_{\text{SSMI}}(\{ {\bf n}_k\})\approx \prod_k Y_{0,0}({\bf n}_k),
\end{equation}
$Y_{0,0}$ is the zeroth spherical Harmonics. $\Psi^e_{\text{SSMI}}$ is 
rotationally
invariant unlike NMI or pBEC; the microscopic wave function 
which yields the desired collective behavior specified in this limit is
(see Fig. 2b))

\begin{equation}
\Psi^e_{\text{SSMI}} \approx \prod_k 
\frac{(\chi^\dagger_{k\alpha}\chi^\dagger_{k\alpha})^n}{\sqrt{(2n+1)!}} |0\rangle .
\end{equation}

The situation for an odd number of atoms per site is more involved
and will be discussed in the next section. 
As mentioned, more detailed discussions on quantum spin ordered or 
disordered phases can be found in two very recent works 
\cite{Demler03,Snoek03}.

\begin{figure}
\begin{center}
\epsfbox{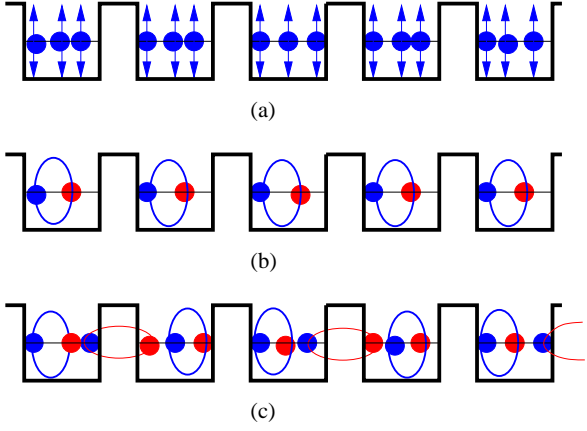}
\leavevmode
\end{center}
\caption{Schematic of
microscopic wave functions of
an NMI (a),  an SSMI for $N_0=2n$ (b) and a DVBC for $N_0=2n+1$ (c).
A dot carrying a double-headed director pointing in direction ${\bf n}$
represents a
spin-one boson in a state
$\psi^{\dagger}_\alpha {\bf n}_\alpha |vac\rangle $; each pair of dark and
light dots connected with a ring
stands for a spin singlet pair of two spin-one atoms.}
\end{figure}

\section{DVBCs in One-Dimensional Mott Insulating States:
small hopping limit}

Consider one-dimensional Mott states when the hopping is much less than $E_{c,s}$.
For an odd number ($N_0=2n+1$) of atoms per site, 
each two unpaired atoms at two neighboring sites form a spin singlet
and the ground state is a dimerized-valence-bond crystal state\cite{Zhou3}.
The previous results suggest the following wave functions (see Fig. 2c))

\begin{equation}
\Psi^o_{\text{SSMI}} \approx \prod'_{k}  
\frac{(\chi^\dagger_{k\alpha}\chi^\dagger_{k\alpha})^{n}
(\chi^\dagger_{(k+1)\alpha}\chi^\dagger_{(k+1)\alpha})^{n}}
{(2n+3)(2n+1)!}
\chi^\dagger_{k\beta}\chi^\dagger_{(k+1)\beta} |0\rangle .
\end{equation}
The product $\prod'$ is carried over all odd sites $(k=2m+1)$ or all 
even ones $(k=2m)$.
Since the key idea of SSQCs is intimately connected with 
properties of DVBCs, in this section
we will present evidence for DVBC states
in one-dimensional lattices.

Though dimerization occurs in all one-dimensional Mott states for an
odd number of particles per site,
we start with a simple limit when 
\begin{equation}
t \ll \sqrt{E_c E_s}, E_c;
\end{equation}
we would like to demonstrate the origin of these 
dimerized states of spin-one atoms from the point of view 
of bilinear-biquadratic (BLBQ) spin 
models. Discussions on more general cases can be found in section 
VIII.
The Hilbert space at each site is defined in Eqs.7-11.
For odd numbers of bosons per site($N_0 \gg 1$),
the low energy Hilbert space for each isolated site is spanned by states of total spin 
$S=1,3,5...$. Therefore,
as $E_s$ is much larger than $t$, one can further truncate the Hilbert space
at each site into a space of a spin-one particle (see FIG.3).

The effective exchange interactions between two adjacent condensates 
can then be obtained by examining the excitation spectrum of 
two coupled 
sites. 
At $t=0$, the ground state of two decoupled condensates has a 
nine-fold degeneracy. 
At any finite $t$ much less than $E_{s,c}$,
because of virtual hopping between two sites, 
the nine-fold degeneracy in the truncated space is lifted and the 
resultant spectrum is

\begin{equation}
E(S)=-\alpha_S J, S=0,1,2;
\label{spectra}
\end{equation}
$\alpha_S$ calculated for two
coupled condensates are

\begin{eqnarray}
&& \alpha_S=\frac{\gamma_1(S)}{1- 4c_s}+\frac{\gamma_2(S)}{1+ 2 c_s}
+\frac{\gamma_3(S)}{1+ 8 c_s},
\label{alpha}
\end{eqnarray}
and

\begin{eqnarray}
&& \gamma_1(0)=\frac{1}{6}, \gamma_3(0)=\frac{2}{15},
\gamma_3(1)=\frac{1}{10}; \nonumber \\
&& \gamma_2(2)=\frac{2}{15}, \gamma_3(2)=\frac{7}{150}; \nonumber \\
&& \gamma_2(0)=\gamma_1(1)=\gamma_2(1)=\gamma_1(2)=0. \nonumber \\
\end{eqnarray}
$c_s=E_s/E_c$;
and $\alpha_0$ is always larger than $\alpha_{1,2}$.

The truncation can be carried out in a similar way in a lattice.
The following reduction in dimensions of the Hilbert spaces takes place
when the hopping is weak

\begin{equation}
{\cal D}_0=[\frac{(N_0+1)(N_0+2)}{2}]^{V_T}
\rightarrow {\cal D}_T=3^{V_T}.
\end{equation}
Here $V_T$ is again the number of sites in a lattice and the number of 
bosons at each side $N_0$ is much larger than one.
In the truncated Hilbert space of dimension ${\cal D}_T$, 
Eq.\ref{spectra} suggests the following effective Hamiltonian

\begin{equation}
{\cal H}_{eff}=-\sum_{\langle kl\rangle }\sum_S J\alpha_S{\cal P}_S({\bf S}_k,{\bf 
S}_l)
\label{eff}
\end{equation}
where ${\cal P}_S({\bf S}_k,{\bf S}_l)$ is a projection operator for
a total spin S state of two condensates at neighboring sites $k$ and 
$l$ (${\bf S}_k^2={\bf S}_l^2=2$).

\begin{figure}
\begin{center}
\epsfbox{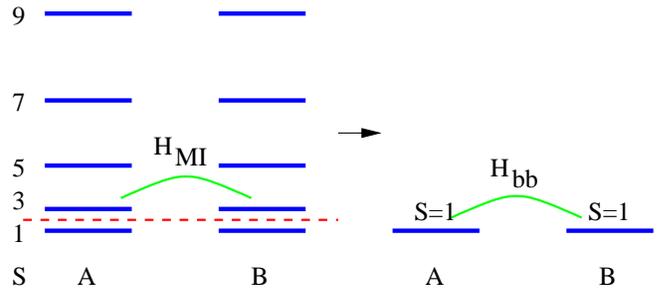}
\leavevmode
\end{center}
\caption{The truncation of the Hilbert space of two condensates A and B
($N_0=2n+1$) when
$t$ is much less than
$E_{c,s}$. The effective Hamiltonian in the truncated space is $\cal{
H}_{\text{b.b.}}$.}
\end{figure}

For one-particle per site ($N_0=1$) at small hopping limit, a
straightforward
calculation yields
values of $\alpha_S$, $S=0,1,2$,

\begin{eqnarray*}
\alpha_1=0,
\alpha_0=\frac{4}{1-2c_s}, \alpha_2=\frac{4}{1+c_s}.
\end{eqnarray*}
and again $\alpha_0 >\alpha_2 >\alpha_1$.

Now we can proceed to
one-dimensional lattices where the situation is relatively well understood.
For an odd number of bosons per site ($N_0=2k+1$), the Mott insulating 
state should be a DVBC with a twofold degeneracy \cite{Zhou3}.
To obtain this result, we take into account
the peculiar property of the truncated Hilbert space and energy 
spectrum. 
Following Eq.\ref{eff}, one shows that the 
problem of interacting spin-one bosons with an odd number of bosons per 
site can be mapped into 
the BLBQ spin-one spin chain model.

\begin{equation}
\frac{{\cal H}_{\text{b.b.}}}{J}=\cos\eta \sum_{\langle ij\rangle }{\bf S}_i \cdot {\bf S}_j 
+ \sin\eta \sum_{\langle ij\rangle }({\bf S}_i \cdot {\bf S}_j)^2,
\label{1D}
\end{equation}
if $\eta$ satisfies
$\tan\eta=
(\alpha_1-\frac{1}{3}\alpha_2-\frac{2}{3}\alpha_0)
(\alpha_1-\alpha_2)^{-1}$
and the sign of $\sin\eta$ is chosen to be
the same as $\alpha_1-\frac{1}{3}\alpha_2-\frac{2}{3}\alpha_0$.
$c_s$ varies from $0$ to $1/4$ when $N_0=2k+1 \gg 1$, and from $0$ to 
$1/2$ when $N_0=1$; consequently, $\eta$ varies as

\begin{equation}
-\frac{\pi}{2} > \eta > \eta_0.
\label{eta} 
\end{equation}
The value of $\eta_0$ in general depends on models used in different 
limits; for the large-$N_0$ case ( $N_0=2k+1 \gg 1$),
$\eta_0=-\frac{\pi}{2}-\arctan \frac{1}{2}$ while for $N_0=1$, $\eta_0 
= -3\pi/4$.
For parameters given in Eq.\ref{eta}, 
following discussions in \cite{Affleck88} we arrive at conclusions that
the ground states should be
dimerized valence bond crystals or DVBCs.
The wave function of DVBCs for $N_0=2k+1$
atoms per site is proposed at the beginning of this section.
At $\eta=0$, the model describes a Heisenburg antiferromagnet;
solutions in this limit were also
studied in \cite{Haldane83}.

Taking into account Eqs.8,31,
one obtains spin-spin correlations in DVBC states.
The DVBC state breaks the crystal translational symmetry and has a
twofold degeneracy; one of degenerate states is characterized by the 
following correlation functions($m\neq n$)

\begin{eqnarray}   
&& \langle {\bf S}_m \cdot {\bf S}_n \rangle \propto
-\frac{1}{8}
\big(1- (-1)^{m}
\big)\big( 1 + (-1)^n \big)
\delta_{n, m+1},
\nonumber \\
&& \langle  {\bf S}_m \cdot {\bf S}_{m+1}
{\bf S}_n \cdot {\bf S}_{n+1} \rangle \propto \frac{1}{16}
\big(1- (-1)^m\big) \big(1- (-1)^n \big).\nonumber \\
\label{correlation}
\end{eqnarray}

Because of the selection rule discussed in Eq.\ref{selec} in section II, 
for even numbers of bosons per site, the Hilbert space is spanned by 
states with total spins of $S=0,2,4..., N_0$.
The lowest energy space can be truncated into a spin singlet state. 
As the coupling $t$ between sites is much smaller than $E_s$, the exchange 
interaction is an 
irrelevant operator;
the ground state remains to be a non-degenerate spin singlet Mott 
insulator as far as
$t$ is much smaller than $E_s$.
In this limit, spin-spin correlations for $m\neq n$ are
zero, i.e.
\begin{equation}
\langle {\bf S}_m\cdot {\bf S}_n\rangle \approx 0.
\end{equation}

\section{Excitations in SSMIs}

In this section, we will demonstrate {\bf charge-e} and
{\bf charge-2e} excitations. The existence of
spinless but charge $e$ ($S=0, Q=1$) excitations
which do not carry full identities of spin one-bosons implies 
the fractionalization of spin-one atoms.
These spinless objects also provide important hints on microscopic 
wave functions of SSQCs discussed in the next section.

\subsection{Charge-e excitations in one-dimensional DVBCs}

In terms of creation operators of a valence bond
at link $\eta$ which connects two neighboring sites $i$ and $j$

\begin{equation}
\phi^{\dagger}_\eta=\frac{1}{\sqrt{3}}({\bf h}^{\dagger}_{i,1}{\bf h}^{\dagger}_{j,-1}+
{\bf h}^{\dagger}_{i,-1}{\bf h}^{\dagger}_{j,1}-{\bf h}^{\dagger}_{i,0}{\bf h}^{\dagger}_{j,0}),
\end{equation}
the wave function of one-dimensional DVBC states can be
written as

\begin{equation}
\Psi_{\text{DVBC}}=\prod'\phi^{\dagger}_\eta.
\end{equation}
The product $\prod'$ is carried over all even or odd links;
${\bf h}^{\dagger}_{i,m}$ is a creation operator of a 
condensate at site 
$i$ with total spin
$S=1$, $S_z=m$.

A one-dimensional DVBC state supports two kinds of elementary 
excitations which are
of topological nature\cite{Zhou3}. A spinless but "charge-e" excitation 
($Q=1, S=0$)
represents a spin singlet state with an extra atom compared with
the Mott ground  state; a chargeless but spinful excitation ($Q=0,S=1$) 
on the other 
hand is a spin-one state but with the same number of atoms as the Mott
state. Both excitations are kink-like and created only by
nonlocal operators.

An $S=1, S_z=m$ excitation can be created by the following product
operator 
\begin{equation}
C^{\dagger}_{\gamma, m}={\cal P}^1_{G.G.} {\bf h}^{\dagger}_{\gamma,m}\prod_{\eta \in 
{\cal C}_{\gamma}}
[\phi_{\eta}^{\dagger} +\phi_{\eta}].
\label{kink}
\end{equation}
The product is carried over all links $\eta$ along 
path ${\cal C}_\gamma$ which starts at site $\gamma$ and ends at  
infinity;
furthermore, ${\cal C}_\gamma$ is chosen such that the first link 
along the path is 
occupied by a valence bond.
${\cal P}^n_{G.G.}$ is a generalized Gutzwiller projection to project out
states with $n$-particles per site. 
Therefore, Eq.\ref{kink} indeed represents
a spin-one domain wall soliton ($m(S_z)=0,\pm 1$) located at site 
$\gamma$ 
in the DVBC state as shown in Fig.4a).  

The hopping matrix of domain-wall 
along an one-dimensional lattice is

\begin{eqnarray*}
{\cal 
T}_{ij}=-\frac{1}{{3}}\alpha_0[\delta_{i,j-2}+\delta_{i,j+2}].
\end{eqnarray*}
Taking it into account, we obtain the following 
band structure for 
spin-one excitations;

\begin{equation}
{E(Q_x)}=J\alpha_0\big(1 -\frac{2}{3}\cos 2 {Q_x} \big)
\label{band}
\end{equation}
with $Q_x$ defined as a crystal momentum of excitations,
$-\pi/2 < Q_x < \pi/2$.
These excitations are purely magnetic and involve no extra bosons;
therefore are "neutral" ($Q=0$).

Besides spin-one "neutral" excitations,
there are also "charged" spinless ($S=0,Q=1$) domain wall excitations. A 
"charged" 
excitation with a 
positive or negative 
charge ($Q =\pm 1$) is 
defined by a creation operator $B^{\dagger}_{\gamma, \pm}$ and

\begin{equation}
C^{\dagger}_{\gamma, m}= \psi_{\gamma,-m} B^{\dagger}_{\gamma,+}  
=\psi^{\dagger}_{\gamma,m} 
B^{\dagger}_{\gamma,-},
\label{frac}
\end{equation} 
$\psi^{\dagger}_{i,m}(\psi_{i,m})$ is a creation (annihilation) operator for a 
spin-one atom with $S_z=m$. 
The situation in one-dimensional optical lattices is therefore similar to
conducting polymers\cite{Heeger88}.

The splitting of the low energy Hilbert space suggests the
fractionalization of atoms. It also suggests a
possible quantum condensate of fractionalized particles of charge-e.
Following Eq.\ref{frac},
as it is added to an one-dimensional 
DVBC, a spin-one atom fractionalizes into
a spin-zero domain wall (with an extra atom thus "charged") and a spin- 
one "neutral" domain wall because of the twofold degeneracy 
for $N_0=2n+1$ atoms per site.

\begin{figure}
\begin{center}
\epsfbox{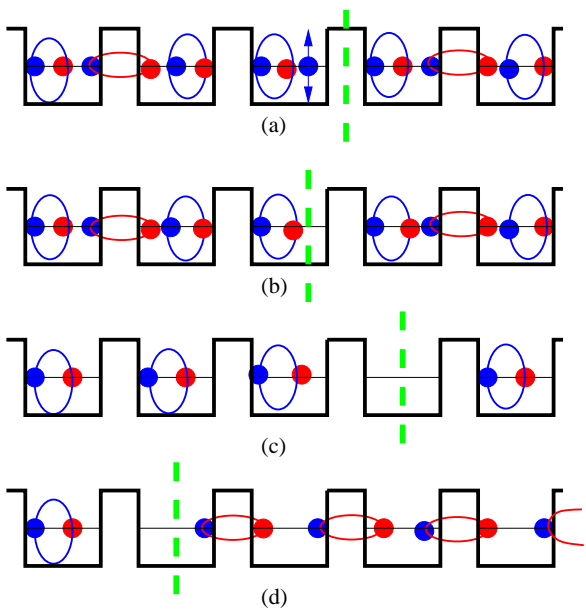}
\leavevmode
\end{center}
\caption{Microscopic wave functions of excitations in one-dimensional
optical lattices.
a) A spin-one kink excitation ($S=1, Q=0$) in a DVBC;
b) A charge-e excitation ($Q=1, S=0$) in a DVBC;
c) A charge-2e
excitation ($Q=2, S=0$) in an
SSMI for an even number of atoms per site. In d), we also show an
$S=0, Q=1$ state in an SSMI ($N_0=2n$);
creation of this state results in a string of singlet bonds
and is an energetic catastrophe.
Locations of excitations are indicated by
dashed lines.}
\end{figure}

\subsection{Charge-2e excitations in one-dimensional SSMIs}

The only spinless excitations in one-dimensional SSMIs for an even number 
of atoms per site
are charge-2e ones.
To demonstrate it microscopically, one considers a limit when $E_s$ is
infinity. All singlet pairs are bound and elementary charge excitations
are charge-2e objects. At large but finite $E_s$, 
one finds at each site, the only spinless excitations are created by 
adding or removing singlet pairs with $Q=2$ and $S=0$.  
These again correspond to charge-2e excitations.

\section{A projected spin singlet Hilbert space based approach} 

To apprehend the key physics in SSMIs and SSQCs, we first provide 
an intuitive approach to this problem.
In a projected Hilbert space,
we are going to demonstrate that the problem of
spin-one bosons with antiferromagnetic interactions
can be mapped into spinless bosons interacting with Ising gauge fields.

Let us consider a projected spin singlet Hilbert space
which satisfies the following conditions:

{\bf a)} each atom has to form a spin singlet with another one either 
at the same site or a nearest neighboring site; 

{\bf b)} each link has to be occupied
by at most one spin singlet pair of atoms (a valence bond).

{\bf c)} Following a), b), the parity of the number of valence bonds 
emitted from
a lattice site is the same as that of the number of particles at that 
site.
For instance, for a site with one particle, one of the links connected to 
that
site has to be occupied by a valence-bond.
For a site with two particles, either two or none of the links are 
occupied by valence-bonds.

In Fig.5, we show examples of states in this projected Hilbert space.
We employ a valence bond counting operator 
\begin{equation}
\hat{d}_{k,k+1}=\frac{1-\sigma^x_{k,k+1}}{2} 
\end{equation}
at each link $\eta=(k, k+1)$,
the eigen values of which are $d_{k,k+1}=0, 1$. So we define 
the projected spin spin singlet Hilbert space as the one 
constructed out of eigenstates of $\{\hat{\rho_k}\}$ and 
$\{\hat{d}_{k,k+1}\}$

\begin{equation}
|...\rho_k, d_{k,k+1}, \rho_{k+1}...\rangle ,
\mbox{ if $(-1)^{\rho_k +\sum_\eta d_\eta}=1$}.
\label{lc} 
\end{equation}
Here $\rho_k$ is the number of particles at site $k$, $d_{k, k+1}$ is the 
number of valence bonds at a link connecting site $k$ and $k+1$.
The local constraint in Eq.\ref{lc} follows point c) discussed above;
$\eta$ represents a link connected to site $k$.

\begin{figure}
\begin{center}
\epsfbox{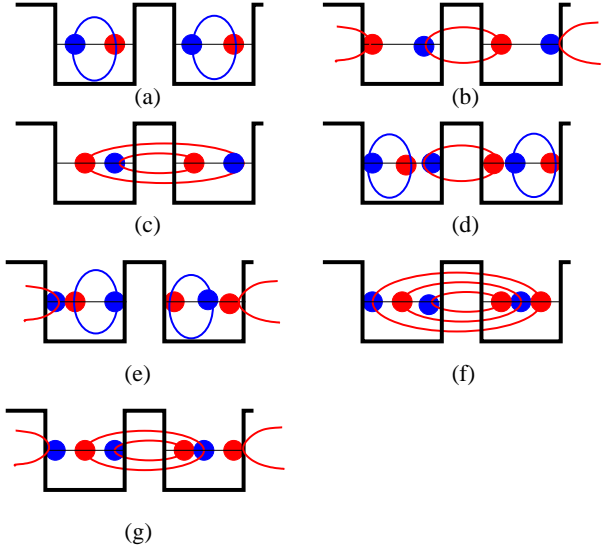}
\leavevmode
\end{center}
\caption{Examples of states in a projected spin singlet Hilbert space.
a), b) and d), e) are states in this space. States in c) or f) and g) can be
"deformed", respectively, into states in a) or d) and e) by locally breaking valence bonds
and constructing intra-site singlets without involving atoms at a third
site.
States in c), f) and g) have higher energies than their counterparts in a), d) and e)
and do not belong to the projected Hilbert space.
In a)-g) dots connected by rings are spin singlet pairs of spin-one
bosons.}
\end{figure}

Consider hopping of an atom between two neighboring sites $k, l$ 
(an even number of atoms per site in initial states, see Fig. 6a).

\begin{eqnarray}
&&  \chi^{\dagger}_{l\alpha}\chi_{k\alpha} 
\frac{(\chi^{\dagger}_{k\alpha_1}\chi^{\dagger}_{k\alpha_1})^n}{\sqrt{(2n+1)!}}
\otimes 
\frac{(\chi^{\dagger}_{l\alpha_2}
\chi^{\dagger}_{l\alpha_2})^n}{\sqrt{(2n+1)!}}
\nonumber \\
&& \rightarrow 
  \chi^{\dagger}_{k\alpha}
\frac{(\chi^{\dagger}_{k\alpha_1}\chi^{\dagger}_{k\alpha_1})^{n-1}}{\sqrt{(2n-1)!}}
\otimes 
\chi^{\dagger}_{l \alpha}
\frac{(\chi^{\dagger}_{l\alpha_2}\chi^{\dagger}_{l\alpha_2})^n}{\sqrt{(2n+1)!}}.
\end{eqnarray}

In a projected
Hilbert space,
hopping of atoms simply leads to transitions between 
the following states $|...\rho_k, d_{k,k+1}, \rho_{k+1}...\rangle $ 

\begin{eqnarray}
&& |...\rho_k,  d_{k,k+1},\rho_{k+1}...\rangle 
\rightarrow \nonumber \\
&& |...\rho_k- 1,1\mp{d}_{k, k+1},\rho_{k+1}+1...\rangle 
\end{eqnarray}
for $d_{k,k+1}=1$ or $0$ in initial states. 
One obtains an effective description of hopping of spin-one atoms
in the projected Hilbert space.

\begin{figure}
\begin{center}
\epsfbox{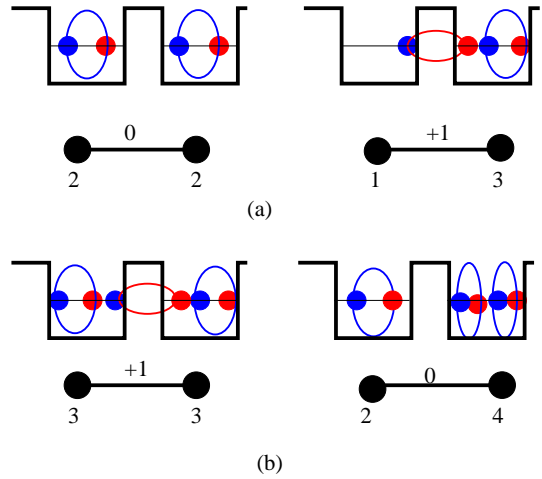}
\leavevmode
\end{center}
\caption{Hopping of atoms in spin singlet Mott states and its
effective description. For an even
number of particles per site in a),
hopping of an atom adds a singlet bond between two sites;
in the lower part of a), this process is represented by changes in
"charges", $b^{\dagger}_kb_k$ at each site (numbers below dots) and
"countings" of valence bonds $\hat{d}_\eta$ at each
link (numbers
above links).
b) is for an odd number of particles per site.
Notice that the parity of $\sum_\eta \hat{d}_\eta+b^{\dagger}_kb_k$ at each site is
conserved when atoms hop around; $\eta$ is a link connected with site $k$.
The description in a) and b) leads to
an effective theory on spin-one atoms in 1D lattices.
In a) and b), we only keep states in the projected Hilbert space
introduced in this subsection.}
\end{figure}

Moreover,
at zero hopping, energies of states with different particles
are given by "charging"
energies in Eq.\ref{hami0}. And to create valence bonds
one needs to break intra site singlet pairs;
and thus the energy cost of a valence bond is $E_s$ if $J_{ex}$ is much smaller 
than $E_s$. 
Therefore in the projected Hilbert space
$|...\rho_k, d_{k,k+1},\rho_{k+1}...\rangle $, we obtain the following mapping

\begin{eqnarray}
&& t \chi^{\dagger}_{k\alpha}\chi_{l\alpha}
\rightarrow t b^{\dagger}_kb_l \sigma^z; \nonumber \\
&&\rho_k \rightarrow b^{\dagger}_kb_k, \frac{\rho_k^2}{2C} \rightarrow
\frac{\rho_{k}^2}{2C}; \nonumber\\
&& \frac{S^2_k}{2I} \rightarrow 2 \Gamma_b \hat{d}_\eta. 
\end{eqnarray}
(using $\sigma^{z}$ as a linear combination of raising and lowering 
operators defined with 
respect to $\hat{d}$).
The resultant Hamiltonian is the same one as in Eq.\ref{LEH}, which will be
derived in a more formal way.
It is easy to confirm that $\Gamma_b \sim 
E_s$ when $E_s \gg J_{ex}$.

\begin{figure}
\begin{center}
\epsfbox{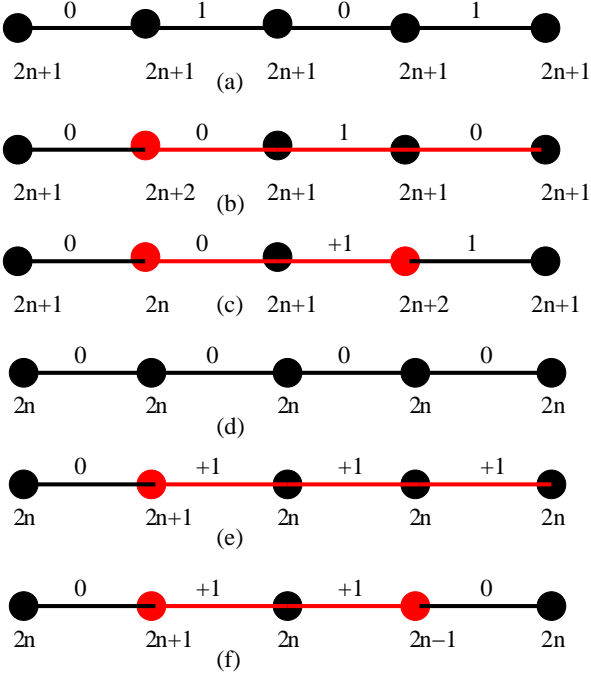}
\leavevmode
\end{center}
\caption{Schematic of one dimensional SSMIs and hopping of bosons in
these states.
a) a DVBC for an odd number of particles per site;
b) an $S=0$ and $Q=1$ kink excitation in a DVBC;
c) hopping of an atom results in a kink-anti kink pair;
d) an SSMIe for an even number of particles per site;
e) an $S=0$ and $Q=1$ "excitation"
in an SSMI (for $N=2n$)
emitting a string of valence bonds;
f) hopping of a boson is suppressed in an SSMI (for $N=2n$) because of
a string of valence bonds between particle-hole excitations.
Numbers appearing at each link are countings of $\hat{d}$ and
those below dots are numbers of particles at each site;
light dots are "charged" and along light links valence bonds have been
removed or added to ground states.}
\end{figure}

The effective Hamiltonian is convenient for the study of spin singlet 
states. 
Let us consider spin singlet Mott insulating states discussed in
section II,III;
In the projected space,
when $E_s \gg J_{ex}$,	
ground states can be read out as

\begin{eqnarray}
&&\rho_k=2n+1, d_{k,k+1}=\frac{1}{2}(1 \pm (-1)^k),\; \mbox{for 
$N_0=2n+1$};\nonumber \\
&&\rho_k=2n, d_{k,k+1}=0,\; \mbox{for $N_0=2n$}
\end{eqnarray}
as illustrated in Fig.7a) and 7d).

The effective Hamiltonian in the projected spin singlet Hilbert space
is also particularly useful for the discussion of hopping of
atoms across lattices.
Following the constraint in Eq.47,
the insertion of an atom to site $k$ changes $\rho_k$ by one unit and
leads to an $S=0, Q=1$ kink-like excitation (see Fig.7b));
this reproduces an spin singlet charge-e
excitation in Fig.4b). On the other hand,
for an even number of particles per site, adding 
an atom to a site results in an $S=0$ and $Q=1$ string-like excitation
(see Fig.7 e)), this was also discussed in Fig.4d).

This difference between lattices with even or odd numbers of 
particles 
per site mentioned above plays a remarkably important role when hopping is 
taken into
account. 
In Fig.7, we illustrate 
hopping of atoms schematically from this point of view.
Hopping of atoms in an "odd" lattice results in kink-anti kink
pairs which are weakly interacting. {\em However, in an "even" lattice
hopping is strongly suppressed because of strings of valence bonds created 
between
particle and hole excitations as shown in Fig.7 f)}; it implies that only 
pairs of atoms
or charge-2e objects propagate along the lattice in this case.
In the following sections, we 
implement this idea more quantitatively using some properties of Ising 
gauge fields.

Similar analysis about one-particle hopping in spin singlet 
states can be easily carried out in high dimensional lattices.
For even numbers of atoms per site, one arrives at the same conclusions
as in one dimensional lattices. For odd numbers of atoms per site,
using the mapping argued in this section, one can also investigate
hopping in possible DVBC states in square lattices (see  
Appendix A).

\section{A fractionalized representation}

For a discussion on SSQCs or more general fractionalized quantum 
condensates (FQCs), we find that it is most 
convenient to introduce
a representation involving "chargeless" ($Q=0$) spin-one operators ${\bf 
a}_{\alpha}$ 
and 
"spinless" ($S=0$) {\bf charge-e}($Q=1$) operators ${b}$;

\begin{eqnarray}
{\bf a}^\dagger_{k\alpha}=b_k {\bf \chi}^\dagger_{k\alpha}.
\end{eqnarray}
${\bf a}$, $b$, $\hat{\rho}$ and ${\bf S}$ satisfy the following
algebras

\begin{eqnarray}
&& [\rho_k, b^\dagger_l]=\delta_{kl} b^\dagger_k, [\hat{\rho}_k, {\bf a}^\dagger_l]=0
\nonumber \\
&& [{\bf S}_k^{\alpha}, {\bf a}^\dagger_{l\beta}]=
i\epsilon^{\alpha\beta\gamma}\delta_{kl} {\bf a}^\dagger_{k\gamma},
[{\bf S}^\alpha_{k}, b^\dagger_l]=0.
\end{eqnarray}
Eq.53 suggests that

\begin{eqnarray*}
\hat{\rho}_k=b^\dagger_kb_k, {\bf S}_{k\alpha}
=-i\epsilon_{\alpha\beta\gamma}{\bf a}^\dagger_{k\beta}{\bf a}_{k\gamma};
\end{eqnarray*}
the spin density is carried by ${\bf a}$-particles, or spin-one 
{\em spinons} 
and charge 
density is carried by $b$-particles, or {\em chargons}. It implies a 
possibility 
of spin-"charge" separation in cold atoms.
General discussions on spin-charge separation in strongly correlated 
electrons can be found in
\cite{Read89,Read91,Senthil99,Senthil00,Rokhsar88,Moessner01,Anderson87,Anderson88,Anderson87a,Affleck88,Ioffe89,Wen89,Wen89a,Lee89,Fradkin90,Read91a,Wen91,Balents98,Sachdev00}.

Obviously, 
the enlarged Hilbert space spanned by $\{\bf a^\dagger\}$ and $\{b^\dagger\}$ includes 
unphysical states violating the symmetry of many-boson wave functions.
Because of the symmetry constraint,
all physical low energy states have to display a superselection 
rule discussed in details in section II.
Namely, for an even number of atoms $N_0$, the low energy Hilbert space 
is occupied by states with 
$S=0,2,4,6..., N_0$ while for an odd $N_0$, the Hilbert space is spanned by 
states with $S=1,3,5...,N_0$.

Following the identity in Eq.10, if one has an even number of ${\bf 
a}^{\dagger}$-particles,
resultant total spins are $S=0,2,4...$ and for an odd number, spin 
states are those with $S=1,3,5...$.
To produce desired symmetric states of spin-one atoms, one has to require 
at least that 
the total numbers of ${\bf a}$- and $b$- particles,
or spinons and chargons, be 
even. 
If the average numbers of ${\bf a}$- and $b$- particles at each site are 
much 
larger than one, it is possible to generate correct low 
energy Hilbert spaces
by imposing the following constraint 
so that for every state $\Psi$,

\begin{eqnarray}
&&\hat{C}_k \Psi=\Psi,
\hat{C}_k=\exp\big(\pm i\pi [{\bf a}^{\dagger}_k\cdot {\bf a}_k + b^{\dagger}_k b_k] 
\big).
\label{constrain}
\end{eqnarray}
The constraint in Eq.\ref{constrain} is precisely
to exclude the unphysical states 
in the enlarged space spanned by $b^{\dagger}$ and ${\bf a}^{\dagger}_\alpha$
which violate the selection rule in Eq.\ref{selec}.

Considerations based on previous works on
discrete gauge symmetries\cite{Zhou01,Demler01} indicate
that the Hamiltonian
in Eq.\ref{hami0} should be equivalent to the following one 
in this fractionalized representation,

\begin{eqnarray}
&& {\cal H}_{FR}=\sum_{k} \frac{{\bf S}_k^2}{2 I} +\frac{{\hat{\rho}}_k^2}{2C} 
-\hat{\rho}_k\mu -\tilde{t} \sum_{\langle kl\rangle }\big({\bf a}^{\dagger}_{k\alpha} {\bf 
a}_{l\alpha}
{b}^{\dagger}_k b_l +\text{h.c.}\big),  
\nonumber \\
\label{Hami}
\end{eqnarray}
In appendix B, we provide further evidence for the equivalence between 
Eq.\ref{Hami} and Eq.\ref{hami0}. 

Let us emphasize that
we are interested in the following limit where the equivalence has been
established;

\begin{equation}
\langle \hat{\rho}_{kb}\rangle =N_0, \langle\hat{\rho}_{k{\bf a}}\rangle =N_{\bf a}
\end{equation}
and both $N_{0}$ and $N_{\bf a}(\ll N_0)$ are much larger than unity. 
\begin{eqnarray*}
\hat{\rho}_{kb}=b^{\dagger}_kb_k, 
\hat{\rho}_{k{\bf a}}={\bf a}^{\dagger}_{k \alpha} {\bf a}_{k 
\alpha};
\end{eqnarray*}
$N_0$ is the number of spin-one particles at each site determined by the 
chemical 
potential; the exact value of $N_{\bf a}$ depends on the trucation of the
low energy Hilbert space and is not important for the rest of discussions.

The resultant Hamiltonian is invariant under a local gauge transformation
\begin{equation}
\hat{C}_k^{-1} {\cal H}_{FR} \hat{C}_k={\cal H}_{FR}.
\label{tran}
\end{equation}

We should also emphasis here that only bilinear operators of
${\bf a}^{\dagger}_k$, $b^{\dagger}$ are invariant under the local gauge transformation 
defined in Eqs.\ref{constrain},\ref{tran},

\begin{eqnarray}
&& \hat{C}_k^{-1} {\bf a}^{\dagger}_{k\alpha}{\bf a}^{\dagger}_{k\beta}
\hat{C}_k={\bf a}^{\dagger}_{k\alpha}{\bf a}^{\dagger}_{k\beta}, \nonumber 
\\
&& \hat{C}_k^{-1} {\bf a}^{\dagger}_{k\alpha}{b}^{\dagger}_{k}
\hat{C}_k={\bf a}^{\dagger}_{k\alpha}{\bf b}^{\dagger}_{k}, \nonumber \\
&& \hat{C}_k^{-1} {b}^{\dagger}_{k}{b}^{\dagger}_{k}
\hat{C}_k={b}^{\dagger}_{k}{b}^{\dagger}_{k}. \nonumber \\
\end{eqnarray}
On the other hand,
all linear operators transform nontrivially under this transformation;

\begin{eqnarray}
&& \hat{C}_k^{-1} {\bf a}^{\dagger}_{k\alpha}
\hat{C}_k=-{\bf a}^{\dagger}_{k\alpha}, \nonumber \\
&& \hat{C}_k^{-1} {b}^{\dagger}_{k}
\hat{C}_k=-{b}^{\dagger}_{k}. \nonumber \\
\end{eqnarray}
From a standard point of view, these linear operators carry charges 
defined with respect to the gauge transformation in Eq.\ref{constrain}
while bilinear operators in Eq.58 are charge neutral.

This property of "fractionalized particles" is more explicit if we 
examine the 
corresponding action of the
Hamiltonian derived in appendix B. 
Besides spinons and chargons, we also have introduced 
discrete gauge
fields $\sigma^x$ defined at links of the lattices
\cite{Wegner71,Wilson74,Fradkin79,Kogut79}.
The constraint discussed above results in
Ising gauge fields ${\sigma^x}_{kl}=\pm 1$ in $D+1$ 
Euclidean space.
And
\begin{eqnarray*}
&& \sigma^z_{kl}=\frac{1}{2i}(\sigma^{\dagger}_{kl}-\sigma^-_{kl}), 
\nonumber \\
&&\frac{1}{2}[\sigma_{kl}^x, \sigma^{\pm }_{k'l'}]=\pm\delta_{kl,k'l'}\sigma^{\pm 
}_{kl}.
\end{eqnarray*}

Following appendix B, an inversion of ${\bf n} \rightarrow -{\bf 
n}$, or $\exp(i\chi) \rightarrow -\exp(i\chi)$ 
corresponds to a discrete gauge transformation 
${\bf a} \rightarrow -{\bf a}$, or
$b \rightarrow -b$.
The action is therefore manifestly invariant under the following
local Ising gauge transformation

\begin{equation}
b_k^{\dagger} \rightarrow \Omega_k b_k^{\dagger},
{\bf a}_k^{\dagger} \rightarrow \Omega_k{\bf a}_k^{\dagger}, 
\sigma^{z}_{kl} \rightarrow 
\Omega_k \sigma^{z}_{kl}
\label{invar}
\end{equation}
with $\Omega_k=\pm 1$.

As indicated here, ${\bf a}^{\dagger}$ and $b^{\dagger}$ are indeed matter 
fields carrying charges of Ising 
gauge fields. Taking into account minimal coupling between 
gauge fields and matter fields, alternatively we 
construct a Hamiltonian which explicitly involves gauge fields 
$\sigma^z_{kl}$, spinon fields ${\bf a}^{\dagger}_k$  and chargon fields $b^{\dagger}_k$.

\begin{eqnarray}
&& {\cal H}_{MG}=\sum_{k} \frac{{\bf S}_k^2}{2 I} 
+\frac{{\rho}_k^2}{2C} 
-\rho_k\mu \nonumber \\
&&
-\tilde{t} 
\sum_{\langle kl\rangle }\sigma^z_{kl}\big({\bf a}^{\dagger}_{k\alpha} 
{\bf 
a}_{l\alpha}+
{b}^{\dagger}_k b_l  +h.c.\big),
\nonumber \\
\label{Hami2}
\end{eqnarray}
The Hilbert space of the Hamiltonian in Eq.\ref{Hami2} is again subject to 
the constraint in Eq.\ref{constrain}. 
The Hamiltonian is
invariant only under the following 
generalized transformation
\begin{eqnarray}
&& \hat{C}'_k=\hat{C}_k \otimes \exp\big(\pm i\pi 
\sum_+\frac{1-\sigma^x_{kl}}{2}\big), \nonumber \\
&& (\hat{C}'_k)^{-1} {\cal H}_{MG} \hat{C}'_k={\cal H}_{MG}.
\label{constrain1}
\end{eqnarray}
The sum over the cross ($+$) is carried over all links connected with site $k$.
Calculations
indeed suggest that the low energy 
physics of the new 
Hamiltonian be the same as the one in Eq.\ref{hami0}\cite{fr}.
Both Eq.\ref{Hami} and Eq.\ref{Hami2} will be employed for
discussions on FQCs and specifically SSQCs.

\section{Spin singlet states in high dimensional lattices}

To facilitate discussions on SSMIs and SSQCs or more general FQCs,
we introduce the following correlation functions

\begin{eqnarray}
&& {\cal G}^{\bf a}_{\alpha\beta}(k,l)=
\langle {\bf a}^{\dagger}_{k\alpha} {\bf 
a}_{l\beta} \rangle _m \nonumber \\
&& {\cal G}^b(k,l)=\langle {b}^{\dagger}_k {b}_l \rangle _m 
\end{eqnarray}
evaluated at a fixed gauge similar to a "minimal gauge" suggested in 
\cite{Fradkin79}.
In this gauge, the number of links where $\sigma_{kl}^z=-1$ is
minimal for a given distribution of gauge fields.
The two-point correlation functions in Eq.64 are expected to diagnose 
possible condensation in the spin or charge sector of the Hilbert space.
In addition, we introduce local pairing
order parameters 

\begin{eqnarray}
&& {\Delta}^{\bf a}_{\alpha\beta}(k,k)=
\langle {\bf a}^{\dagger}_{k\alpha}{\bf a}^{\dagger}_{k\beta}  \rangle 
\nonumber \\
&& {\Delta}^b(k,k)=\langle  b^{\dagger}_k b^{\dagger}_k  \rangle 
\end{eqnarray}
which
are manifestly gauge invariant under the local 
gauge transformation defined in Eq.\ref{tran}.

We further assume that the two parameters 

\begin{eqnarray*}
G_s=\frac{t}{E_s}, G_c=\frac{t}{E_c}
\end{eqnarray*}
can be varied independently. 
Two critical values $G_{sc}, G_{cc}$ are
introduced below to facilitate discussions on SSQCs and SSMIs. 
For the purpose of demonstration, in this section we limit ourselves 
to lattices
with an even number of bosons per site($N_0=2n$), though
a generalization using the scheme developed in section V is 
possible.

{\bf Large Hopping limit}: $G_s > G_{sc}$, $G_c > G_{cc}$
Under this condition, 
both spinons and chargons 
condense 
and the ground state
represents a {\em polar condensate} or pBEC. 
The gauge fields are screened and $\prod_{\Box}\sigma^z=1$ 
at each plaquette; here $\Box$ represents an elementary plaquette.
There are two branches
massless spin wave excitations in this rotational- 
symmetry-breaking state and one branch massless density mode.
The ac Josephson oscillation frequency is precisely the difference between 
the chemical potentials of two condensates.

{\bf Intermediate Hopping limit I}: $G_s > G_{sc}$, $G_c < G_{cc}$
When $E_s$ is much less than $E_c$, this corresponds 
to a limit
where $E_c G_{cc} > t > E_s G_{sc}$.
Following the Hamiltonian, chargons are completely depleted 
from the condensate. 
We are particularly interested in a self-consistent solution where 
spin-one spinons "condense" one way or the other.
States in this limits are incompressible and therefore naturally 
correspond to  
Mott insulating states; condensation of spinons leads to 
nematic long range order. 
Nematic Mott states were studied briefly in section II; in appendix C,
we provide an alternative description on NMIs using this fractionalized
representation.

{\bf Intermediate Hopping limit II}: 
$G_s < G_{sc}$, $G_c > G_{cc}$
This limit can be relevant to atomic gases in high 
dimensional optical lattices, especially
when $E_s$ is not too small compared with $E_c$.
We are going to present results in this case,
for the completeness of analysises on SSQCs and more important
to facilitate discussions on low dimensional optical lattices in the next 
section.
This case represents a situation
where $E_s G_{sc}> t > E_c G_{cc}$. 

Discussions in this case can be carried out parallel to the previous 
case analyzed in appendix C.
All spinful excitations are fully gapped; 
after integrating out these excitations, we obtain 
an effective Hamiltonian in terms of chargons

\begin{eqnarray}
&& {\cal H}_{fqcb}=
\sum_k \frac{{\hat{\rho}}^2_k}{2C}-\mu\hat{\rho}_k 
-\tilde{t}\sum_{\langle kl\rangle }({b}^{\dagger}_{k}{b}_{l}
\sigma^z_{kl}+h.c.)\nonumber \\
&& +\Gamma_b \sum_{\langle kl\rangle }\sigma^x_{kl}-K_{sb} 
\sum_{\Box}\prod_{\Box}\sigma^z_{kl}.
\label{fqcb}
\end{eqnarray}
$\Gamma_b, K_{sb}$ are again functions of $t, E_s$.
$K_{sb}$ is much less than $\Gamma_b$ 
when $G_s$ is small, 
\begin{equation}
\frac{K_{sb}}{\Gamma_b}\ll  1 (\Gamma_b\sim 6E_s);
\end{equation}
and expected to 
be divergent close to $G_{sc}$.
The induced constraint on the Hilbert space is

\begin{eqnarray}
\hat{C}_k^{fb} \Psi=\Psi, \hat{C}_k^{fb}=\exp\big(i\pi [{b}^{\dagger}_k{b}_k
+\sum_{+}\frac{1-\sigma^x_{kl}}{2}]\big).
\label{constraintfb}
\end{eqnarray}
And the Hamiltonian is also locally invariant under the action of
$\hat{C}_k^{fb}$,

\begin{equation}
\big(\hat{C}_k^{fb}\big)^{-1} {\cal H}_{fqcb} \hat{C}_k^{fb}=
{\cal H}_{fqcb}.
\end{equation}
If $\Gamma_b$ is much less than $K_{sb}$, chargons condense and the 
phase coherence is maintained, 

\begin{equation}
{\cal G}^b(k,k+\infty)\neq 0,
{\cal G}^{\bf a}_{\alpha\beta}(k,k+\infty)=0
\end{equation}
and the rotational symmetry is 
unbroken.
This state is compressible but should have zero spin susceptibility at 
a low temperature limit.
We define this state as a charge-e SSQC.

One can also carry out discussions on a limit
when $t$ is much less $E_s$, or
\begin{equation}
\frac{\Gamma_b}{K_{sb}} \gg 1.
\end{equation}
And gauge fields $\sigma^x$ are expected to
be unity at each link over the whole lattice; under 
the constraint 
in Eq.\ref{constrain}, the reduced Hilbert 
space is subject to a constraint

\begin{equation}
\hat{\cal C}^b_k \Psi=\Psi,
\hat{\cal C}^b_k =\exp(i\pi b^{\dagger}_kb_k).
\end{equation}
Thus, the low energy physics is 
determined by the 
pair hopping of chargons;

\begin{eqnarray}
&& {\cal H}_{pb}=
\sum_k \frac{\rho^2_k}{2C} -\mu\rho_k
-\tilde{t}_2\sum_{kl}({b}^{\dagger}_{k}
{b}^{\dagger}_{k}{b}_{l}{b}_{l}+h.c.),
\nonumber \\
&&( {\hat{C}^b_k})^{-1} {\cal H}_{pb} {\hat{C}^b_k}={\cal H}_{pb}.
\label{pb} 
\end{eqnarray}
and

\begin{eqnarray}
&& \tilde{t}_2\approx \frac{t^2}{\Gamma_b} (\Gamma_b\sim 6E_s).
\end{eqnarray}
When $t_2= N_0^2 \tilde{t}_2 \gg E_c$, 
the ground state is a condensate of paired chargons; 
in terms of chargon-operator $b^{\dagger}$,

\begin{eqnarray}
&& |g_2\rangle =\prod_k {\cal P}_{N_0\times V_T}
\exp\big( \Phi_b^{\dagger}(0)\big) |0\rangle ,\nonumber \\
&& \Phi^{\dagger}_b(Q_0)=\Delta_0 
\sum_q b^{\dagger}_{Q_0+q}b^{\dagger}_{-q}.
\label{pFQCb}
\end{eqnarray}
The projection operator ${\cal P}_{N_0\times V_T}$ acting on a 
generalized BCS pairing 
wave function
is to project out $N_0V_T$-particle states; $N_0$ and $N_a$ are  
even numbers.
Again,
${\cal G}^{\bf a}_{\alpha\beta}(k,l)$ and
${\cal G}^b(k,l)$ vanish as $k-l$ approaches infinity but
Eq.\ref{pFQCb} indicates

\begin{equation}
\Delta^b(k,k)=\frac{\Delta_0}{1-\Delta_0^2} 
\end{equation}
in the ground state.
Finally, we should emphasize that this charge-2e SSQC
could be alternatively considered as condensation of "charged" 
soliton-anti soliton pairs from a point of view of section IV.

{\bf Small hopping limit: $G_s < G_{sc}$, $G_c < G_{cc}$}

We are going to revisit the strong-coupling limit of
spin-one atoms in Mott insulating states; we will employ the 
fractionalized representation to express the wave functions of 
SSMI. Discussions are limited to even numbers of bosons per site.

In a strong-coupling limit $G_s, G_c \ll 
1$ and $t_1 \ll E_s$, $t_2\ll E_c$, gauge fields of $\sigma^x$ 
"condense".
The Hilbert space of spinons and chargons explicitly satisfies a 
constraint $\hat{\cal C}^{\bf a}_{k}\hat{\cal C}^b_k \Psi=\Psi$.
($\hat{\cal C}^{\bf a}_{k}$ is defined in Appendix C.)

As $t$ is much smaller than the interaction energies,
the Hamiltonian commutes with local spin and number operators
in the leading order of $E_{c,s}$;

\begin{equation}
[{\cal H}, {\bf S}_k^2]=[{\cal H}, {\rho}^2_k]\approx 0.
\end{equation}

Therefore, the ground state should be an eigenstate of these two {\em 
local} operators
and the wave function can be expressed as a projected BCS state of 
chargons

\begin{equation}
|g_3\rangle ={\cal P}^{2n}_{G.G.}|g_2\rangle 
\label{MI}
\end{equation}
where ${\cal P}^{2n}_{G.G.}$ is a generalized Gutzwiller projection 
for $2n$-chargons per site.
One can show that $S^2_k=0$ for any $k$ and Eq.\ref{MI} represents a spin 
singlet Mott insulator (SSMI) studied in section II.

Before turning to SSQCs in one-dimensional lattices, 
we first revisit SSMIs in one-dimensional lattices to further 
understand the even-odd effect discussed in section III. 
We are going to generalize previous results on one- 
dimensional SSMIs using mappings to an instanton
gas model of Ising gauge fields and quantum dimer models.

\section{SSMIs in low dimensional lattices: A General approach}

\subsection{Instantons in Ising gauge fields}

From the point of view of Ising gauge fields introduced in the previous 
section, distinct properties of Mott states 
for odd and even numbers of bosons per site discussed in section III can 
also be attributed to a topological term in the effective actions for even and 
odd numbers of bosons per site. 
In this section, we are going to revisit SSMIs in low dimensional 
lattices employing the effective descriptions introduced in section VI. 
We will reproduce results in section II and III; furthermore, in 
one-dimensional lattices 
we are 
going to generalize the previous results on DVBC states in the 
limit $t \ll \sqrt{E_s E_c}$ 
(see section III) to the entire Mott phase.

The quantum problem of the Hamiltonian ${\cal H}_{FR}$, or equivalently
the Hamiltonian of constrained quantum rotor model ${\cal H}_{CQR}$
defined in
$d$ dimensions can
be mapped into the
following
model in $(d+1)$ dimensions (appendix B),

\begin{equation}
{\cal H}_{FR} \leftarrow\rightarrow {\cal H}_{CQR} \rightarrow 
O(2)\otimes O(3)\otimes Z_2. 
\end{equation}
The $O(2)\otimes O(3)$ non-linear sigma model
characterizes the dynamics of two variables ${\bf n}, \exp(i\chi)$
introduced in section II, or spinons ${\bf a}^+$ and chargons $b^+$ in 
discussed in section VI.

In high dimensional lattices when $t$ is much less than $E_s$ and $E_c$,
spinons don't condense and chargons are localized because of repulsive 
interactions; and
the director ${\bf n}$ becomes disordered as suggested by a renormalization group
equation analysis on the $O(3)$ nonlinear sigma model.
For an even number of atoms per site, this strong coupling fixed
point can be confirmed by directly solving the Hamiltonian ${\cal H}_{MI}$
in the same limit in all dimensions.
In one-dimensional lattices, spin-one spinons again do not condense and 
the director ${\bf n}$ is quantum disordered 
at any finite $E_s$ because of long range fluctuations.

For an odd number of atoms per site in the absence of hopping, 
the low energy space is highly degenerate (see section III)
and the situation is more delicated. 
On the other hand, from the point of view of effective actions, the 
difference
in numbers of atoms per site is only reflected in a Berry's phase
term(see Appendix B).
As far as low dimensional lattices are concerned,
we speculate that 
this topological term doesn't alter the 
renormalization group flow of the $O(3)$-model and arrive at the conclusion 
that 
for an odd 
number of atoms per site, spinons also do not condense and the director 
${\bf n}$ is also disordered as the 
strong-coupling limit is approached\cite{num}. This observation at least 
can be 
justified in 
one-dimensional lattices by comparing the results obtained below and the 
ones presented in section III.

In this paper,
we would like to assume
a quantum disordered phase in square lattices for odd numbers of atoms per 
site because of certain frustration. For discussions on square 
lattices with odd numbers of atoms per site,
we will take this as a starting point and analyze 
disordered phases in section VIII B3.

In the strong coupling limit where the hopping integral is much less than 
$E_{s,c}$, 
after integrating over $O(2)$ and $O(3)$ degrees of freedom, or chargons 
and spinons, one 
obtains an effective low energy theory,

\begin{eqnarray}
&& {\cal S}=-k_\tau\sum^\tau_{\Box}\prod_{\Box}\sigma^z_{kl} 
-k_s\sum_{\Box}^s\prod_{\Box}\sigma^z_{kl}\nonumber \\
&& + i N_0\frac{\pi}{2}\sum_{r}(1-\sigma_{k k+\hat{\epsilon}}^z).
\label{tt}
\end{eqnarray}
$\sigma^z_{kl}$ is an Ising field defined at each link $kl$;
$\Box$ is an elementary plaquette in a $(1+d)$D Euclidean space.
$\sum^s$ and $\sum^\tau$ are carried over spatial plaquettes and  
plaquettes
involving temporal links respectively. 
$k_{s,\tau}$ can be estimated using a high temperature expansion

\begin{equation}
k_s\sim (\epsilon t)^4,
k_\tau \sim \mbox{max} \big( \frac{t^2}{E_c^2}, \frac{t^2}{E_s^2} \big).
\label{K}
\end{equation}
$\epsilon^{-1}$ is a high energy cut-off in our problem.

In one-dimensional lattices, the term proportional to $k_s$ is absent;
furthermore, the result in Eq.\ref{tt} is valid as far as 

\begin{eqnarray*}
t < G_{cc} E_c.
\end{eqnarray*}
And therefore conclusions on one-dimensional lattices arrived in this 
section are expected to be valid for all one-dimensional Mott states 
disregarding
the ratio between $t$ and $E_s$.

It is worth emphasising that only the topological term explicitly depends on
numbers of particles at each site. 
General effects of Berry's phases on spin liquids have been 
studied in various papers on strongly correlated electrons
\cite{Haldane83,Haldane88,Read89,Read91,Jalabert91,Senthil99,Moessner01}.
The topological term in Eq.\ref{tt} indeed has rather surprising 
consequencies on 
many-body ground state degeneracies and quantum numbers of excitations. 
In fact, this term determines the confining-deconfining property of  
effective Ising gauge fields in one-dimensional lattices. It can be easily 
appreciated in terms of 
suppression 
of instantons. 

Consider a $Z_2$-instanton in $(1+1)d$ Euclidean space which emits a 
string of 
negative bonds pierced by the $x$-axis so that the boundary condition along 
the temporal direction is periodic; only at the $X_{th}$ plaquette shown 
in Fig.8,

\begin{equation}
\prod_{\Box}\sigma^z_{ij}=-1;
\end{equation}
otherwise the plaquette integral over an elementary square is one.

\begin{figure}
\begin{center}
\epsfbox{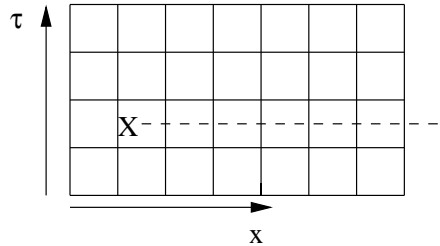}
\leavevmode
\end{center}
\caption{A $Z_2$-instanton located at $X_{th}$ plaquette in $(1+1)$D. At
links
pierced by the dashed line,
$\sigma_{ij}^z=-1$.}
\end{figure}

Following the action in Eq.\ref{tt}, an instanton located at $X$th
plaquette
along the $x$-axis has a finite action and carries a phase

\begin{equation}
\Gamma_B=X \pi N_0.
\end{equation}
In a dilute instanton gas limit (o( $e^{-k_\tau}$)), the partition 
function of instantons is

\begin{eqnarray}
&&{\cal Z}_{DIG}=\sum_{n}\frac{{\beta}^n}{n!}\sum_{\{X_1\}}
...\sum_{\{X_n\}}
\nonumber \\
&& \exp[-n k_\tau+ i\pi \sum_{m=1,}^{n} X_m N_0 ]
\end{eqnarray}
where the subscript $m$ labels instantons.
For $N_0=2n+1$,
the sign of the amplitude of an instanton therefore 
alternates between positive and negative ones as the location of the 
instanton 
is shifted by one lattice constant $a$ along the $x$-axis.
A direct evaluation shows that in this case in the leading order of 
$o(e^{-k_\tau})$,

\begin{equation}
\frac{\ln {\cal Z}_{DIG}}{\beta L_x} \rightarrow 0
\end{equation}
as the perimeters along $x$ and $\tau$ direction in the Euclidean space go 
to infinity.
It suggests that instantons of $Z_2$ type should be absent
in $(1+1)$D because of destructive interferences\cite{Haldane02}. 

The alternating sign of amplitude also implies
that the ground state should break the 
crystal translational symmetry.
Furthermore, the amplitude is periodical
as a function of centers of instantons 
with a period of $2a$. So the ground 
state should develop a spin-Peirls structure with a period doubling 
that of
underlying lattices.
The suppression of instantons therefore appears to lead to unusual 
deconfining 
Ising gauge fields in 
$(1+1)$D and 
a twofold degenerate spin-Peirls state, to which DVBCs belong.
For an even number $N_0$, obviously instantons proliferate in the 
system and the ground state is non-degenerate and translationally 
invariant.

The instantons' behavior in this limit also determines quantum numbers 
of low lying {\em elementary} excitations, especially "topological" ones.
For an odd number of bosons per site, a kink shown in Fig.4
should be an elementary spin-one excitation in the spectrum; this 
is indicated by the deconfining property of (frustrated) Ising gauge 
fields 
in (1+1)D. 
Thus, for an odd number of atoms per site, spin-charge
separation takes place in any one-dimensional Mott state. 
On the other hand,
for even numbers of atoms per site, gauge fields are confining. 
Lowest lying elementary excitations
carry $S=1$, $Q=1$, or $Q=0, S=2$ or $S=0$, $Q=2$.
The first one represents adding one atom to the system;
the second type of excitations is to flip one spin 
singlet into $S=2$ state and the last one is to add (or remove) a singlet 
pair of two spin one bosons (Fig.4c).

Generally,
topological terms are known to play very important roles in quantum 
disordered
states. In one-dimensional Heisenberg antiferromagnetic spin chains, 
topological
terms lead to gapless spin liquids in half-integer spin chains, differing 
from an integer spin chain\cite{Haldane83}. In 2D, 
topological terms can
result in spin-Peirls states with different periodicity for half integer,
odd and even integer Heisenberg antiferromagnets 
(disordered states)\cite{Haldane88,Read89}.
In the current situation the topological terms 
determine distinct quantum
numbers of excitations for even or odd numbers of atoms per site.

\subsection{Quantum dimer models}

The Ising gauge fields describe the 
dynamics of low lying 
collective spin states of condensates at each site. 
Let us consider

\begin{eqnarray}
&& \Sigma_{ij}^x =\phi^{\dagger}_{ij}\phi_{ij}-\frac{1}{2},
\nonumber \\
&& \Sigma_{ij}^y=\frac{1}{{2}}(\phi^{\dagger}_{ij}+\phi_{ij}),
\nonumber \\
&& \Sigma_{ij}^z=\frac{1}{{2}i}(\phi^{\dagger}_{ij}-\phi_{ij}).
\nonumber \\
&& \phi^{\dagger}_{ij}=\frac{1}{\sqrt{3}} ({\bf h}^{\dagger}_{i,-1} {\bf h}^{\dagger}_{j,1}
+{\bf h}^{\dagger}_{i,1}{\bf h}^{\dagger}_{j, -1}-{\bf h}^{\dagger}_{i,0}{\bf h}^{\dagger}_{j,0})
\label{creation}
\end{eqnarray}
where $\phi^{\dagger}_{ij}$ is the creation operator of a singlet state of two 
adjacent condensates, or a valence bond;
${\bf h}^{\dagger}_{m_F}$ is a creation operator of a condensate 
with total spin one.

In a projected space of valence bond 
configurations where each link is occupied by at most one valence 
bond defined by $\phi^{\dagger}_{ij}$, the Hilbert space at each link is an Ising 
doublet as pointed out in section V.
One can easily confirm that $\Sigma_{ij}^\alpha$, $\alpha=x,y,z$ satisfy 
the algebra of $\sigma_{ij}^\alpha$ and can be identified as
$\sigma_{ij}^\alpha$.
So one interprets the effective action 
in terms of dynamics of valence bonds.

An interesting alternative but closely related point of view is the 
quantum dimer model\cite{Rokhsar88}. 
The effective Hamiltonian of the Ising gauge fields in 
Eq.\ref{tt} 
can be written as
\begin{equation}
{\cal H}_{IG}=-\Gamma \sum_{\langle kl\rangle }\sigma_{kl}^x -K_s\sum_\Box \prod_\Box 
\sigma^z
\label{IG}
\end{equation}
where the first sum is over all links and the second sum is over spatial 
plaquettes; and in one-dimensional lattices, the second 
term 
is absent.
One can verify that the action of ${\cal H}_{IG}$ is precisely
that of the Ising gauge fields in Eq.\ref{tt}\cite{Kogut79};
indeed, 

\begin{eqnarray}
&& k_s=\epsilon K_s,
k_\tau = -\frac{1}{2} \ln \tanh (\epsilon \Gamma ).
\end{eqnarray}
When $\Gamma$ is much smaller than $K_s$, 
$k_\tau$ becomes much larger than unity.
On the other hand, $k_\tau$ is much less than the unity as $\Gamma$ 
becomes
much larger than $K_s$.

To generate desired topological terms in the action, we require that
the Hilbert space satisfy the following local invariance at each site

\begin{equation}
\hat{g}_k \Psi =\Psi, 
\hat{g}_k=\exp[i\pi(\sum_{+}\frac{1-\sigma^x_{kl}}{2} +N_0)],
\label{cqd}
\end{equation}
for an even ($N_0=2n$) or odd number
($N_0=2n+1$) of atoms each site.
This constraint is identical to the constraint in Eq.48 which was derived
from a very different consideration.
The sum $+$ is carried over all links connected with site $k$.
One can easily show that again for each site $k$

\begin{equation}
\hat{g_k}^{-1}{\cal H}_{IG} \hat{g_k}={\cal H}_{IG}.
\end{equation}
or ${\cal H}_{IG}$ is locally invariant under the action of $\hat{g}_k$.
The parity of the number of bosons enters the theory only in the definition 
of the Hilbert space.

At each link lives a two-dimensional Hilbert space 
$\sigma^x=\pm 1$. By assuming each link either occupied by  
a dimer ($\sigma^x=-1$) or empty ($\sigma^x=1$), one introduces 

\begin{equation}
\hat{d}=\frac{1-\sigma^x}{2}
\end{equation}
as a dimer counting operator (see also section V). 
The first term in ${\cal H}_{IG}$ can be interpreted as the chemical 
potential of dimers.
As $\sigma^z$ is a linear combination of creation and annihilation 
operators of dimers, in square lattices
the last term is the "kinetic" energy of dimers,
which conserves the parity of numbers of dimers emitted from each site
but does not conserve numbers of dimers.
The connection between Ising gauge fields 
and quantum dimer model was recently illustrated 
in an interesting work\cite{Moessner01}.

The Hilbert space has four distinct sectors
if periodical boundaries are imposed along both $x$ and $y$-directions.
This aspect of the generalized model can be demonstrated explicitly in a
torus or ring geometry. Let us introduce winding number operators

\begin{equation}
\hat{T}_{x,y}=\prod_{{\cal C}^{x,y}_\infty} (\sigma^x), \hat{T}_{x,y}^2=1.
\end{equation}
Here ${\cal C}^{x(y)}_\infty$ is a path extending from $-\infty$ to 
$+\infty$ 
along the $x$($y$)-direction;
the product is carried over all 
vertical links ($y$-direction)
or horizontal links ($x$-direction)
pierced by the path. 
Similar global operators were previously introduced for the study of 
correlated states\cite{Senthil00,Moessner01}.

It is easy to verify that for $i=x,y$,

\begin{equation}
[\hat{T}_{i}, \hat{T}_j]=0,
[\hat{T}_{i}, {\cal H}_{IG}]=0.
\end{equation}
$\hat{T}_{i}=\pm 1$ are good quantum numbers and
the Hilbert space therefore
has four sectors with different parities defined by
$\hat{T}_{x,y}$.
For a cylinder or a ring, one can only define $\hat{T}_y$ if $x$ is the
direction of circumference.  
In this case, there are two sectors in the Hilbert space.

\subsubsection{Even numbers of atoms per site in all bipartite lattices}

Let us now turn to the ground states of the Hamiltonian in Eq.\ref{IG}.
Solutions depend on the parity of numbers of bosons per site and we start
with an even number of bosons per site. Following Eq.\ref{K}, as $t$ goes
to zero, $k_{s,\tau}$ approach zero; therefore in this limit,
\begin{equation} 
\frac{\Gamma}{K_s} \gg 1. 
\end{equation} 
The relevant Hamiltonian is 
\begin{equation} 
{\cal H}_{qdm}^{e}=2
\Gamma \sum_{\eta} \hat{d}_{\eta} 
\end{equation} 
which commutes with the
dimer counting operator $\hat{d}_\eta$ defined at each link $\eta$. The 
Hamiltonian is
positive defined and the energy has a lower bound of zero. The ground
state $|G^e\rangle $ obviously is a vacuum of dimers with ${\hat d}=0$ at each
link and is nondegenerate. In one-dimensional lattices, the ground state 
is

\begin{equation}
\hat{d}_\eta |G^e\rangle =0, \hat{T}_y|G^e\rangle =|G^e\rangle .
\end{equation}
where $\eta$ is an arbitrary link.

In a torus, it has an even parity of $\hat{T}_{x,y}$; 
that is
\begin{equation}
\hat{d}_\eta |G^e\rangle =0, \hat{T}_{x,y} |G^e\rangle =|G^e\rangle .
\end{equation}
We identify this state as an SSMI state obtained in section II.
The lowest excitations in this sector
have energies $4\Gamma$ higher than the ground state. 
One can also show that
the lowest energy states in $\hat{T}_x=-1$ or $\hat{T}_y=-1$ sector 
have energies proportional to the system size;
they are not degenerate with 
the global ground state.

\subsubsection{Odd numbers of atoms per site in 1D lattices}

When $t$ is much less than $E_{s,c}$
for an odd number $N_0$, 
the low energy Hilbert space becomes spanned by configurations where
only one of all links connected with each site is occupied by a dimer.
In 1D, the second kinetic term in Eq.\ref{IG} is absent and 
the constraint leads to a dimer crystal 
ground state with a twofold
degeneracy. 
Particularly, two ground states $|G^o_{1,2}\rangle $ have different 
$\hat{T}_y$-parities

\begin{equation}
\hat{d}_\eta |G^o_{1,2}\rangle =\frac{1\pm (-1)^{{m}_\eta}}{2}|G^o_{1,2}\rangle ,
\hat{T}_y|G^o_{1,2}\rangle =\pm |G^o_{1,2}\rangle .
\end{equation}
Here $m_\eta$ numerates link $\eta$ in one-dimensional lattices.
Interpreting each dimer as a valence bond, this 
point of view precisely leads to a dimerized-valence-bond crystal state.

\subsubsection{Odd numbers of atoms per site in square lattices}

For quantum spin disordered states
in square lattices, one arrives at the following conclusions.
For an odd number of atoms, 
the number of dimers from each site should be $1,3,5...$ etc;
however,
as $\Gamma$ goes to infinity, 
the low energy Hilbert space is spanned by configurations with 
one 
dimer 
emitted from each 
site. Consequently, the constraint in this 
limit conserves the number of dimers emitted from each site
and results in the following identity for any $U(1)$ gauge 
choice $\phi$,

\begin{equation}
\hat{h}_k=\exp[i\phi (\sum_{+} \hat{d}_{\eta}-1)], \hat{h}_k \Psi=\Psi.
\label{U(1)}
\end{equation}
The mapping between a number conserved dimer model
and a compact $U(1)$ theory was obtained in \cite{Fradkin90}.

In square lattices, the constraint for $N_0=2n+1$
atoms per site therefore
still leaves an exponentially large degeneracy in the low energy 
manifold; correspondingly,
the kinetic energy in the Hamiltonian 
becomes

\begin{equation}
{\cal H}^o_{qdm}=-K_s\sum_\Box \sigma^{\dagger}\sigma^-\sigma^{\dagger}\sigma^- +h.c.
\end{equation}
in the reduced Hilbert space. 
This term lifts the residual degeneracy in the low energy manifold
($\sigma^{\pm}$ are the raising and lowering operators
defined in $\sigma^x$ basis.); on the other hand,
it preserves the reduced Hilbert space 

\begin{equation}
\hat{h}^{-1}_k {\cal H}^o_{qdm}\hat{h}_k={\cal H}^o_{qdm}.
\end{equation}
Terms such as $\sigma^{\dagger}\sigma^{\dagger}\sigma^{\dagger}\sigma^-$, or $\sigma^-\sigma^-
\sigma^-\sigma^{\dagger}$ are not invariant under the local $U(1)$ gauge 
transformation and have zero matrix elements in the space defined in Eq.
\ref{U(1)}.

It can be shown that the Hamiltonian acting on a reduced Hilbert space defined in Eq.\ref{U(1)}
is equivalent to
the Rokhsar-Kivelson's quantum dimer 
model with $V=0$\cite{Moessner01,Rokhsar88}.
The ground states in a square lattice at $V=0$ are
column states
with a fourfold degeneracy; furthermore, they break the crystal 
translational 
symmetry. This implies that
a quantum disordered state in a square lattice should be a dimerized 
valence bond 
crystal state.
For discussions on excitations, see appendix A.

To summarize, 
two main approaches have been employed to study spin correlated Mott 
insulating physics.
The first one is using the effective Hamiltonian in reduced Hilbert spaces.
The problem of interacting spin-one bosons is first mapped into a 
constrained quantum rotor model; spin 
correlations in Mott states then are 
studied using the Hamiltonian ${\cal H}_{MI}$ introduced in 
section II. In the strong coupling limit when $J_{ex}$ is much less than 
$E_s$, for an odd number of atoms per site,
the Hamiltonian is further equivalent to the bilinear-biquadratic Hamiltonian
${\cal H}_{b.b.}$.

The second approach is to map the quantum problem into a classical one;
the CQR problem can be mapped into 
a classical $O(2)\otimes O(3)\otimes Z_2$ model.
In Mott states, the corresponding classical model is
an $O(3)\otimes Z_2$ one, with a topological term.  
At the strong coupling limit, the model is also equivalent to 
constrained quantum dimer models.
This mapping is employed to investigate disordered states in both one- 
dimensional and 
square lattices.
For an odd number of particles per site,
the later approach is used to generalize the results derived in the first 
approach to an intermediate hopping limit where the first approach is
invalid.

At last,
in square lattices, the effect of topological terms 
when $N_0$ is odd can be 
conveniently studied by a duality transformation from a constrained Ising 
gauge model to
a fully frustrated Ising model in a transverse 
field\cite{Jalabert91,Senthil99,Moessner01,Sachdev00}. 
The duality transformation indicates the ground state at 
infinit $\Gamma$ limit should break the crystal translational 
symmetry. This again is consistent with the point of view of quantum dimer
models.

\section{Evidence for SSQCs in one-dimensional optical lattices}

Following the analysises in section VI, it is clear that to have  
SSQCs in high dimensional lattices for integer numbers of particles per site, 
$E_s$ has to be at least comparable to $E_c$. 
On the other hand, following analysises in \cite{Zhou01},
because of long wave length fluctuations, 
in one-dimensional lattices spin correlations are always short 
ranged; the rotational symmetry is unbroken (see Fig.9). 
Spin excitations are fully gapped for any finite value of $E_s$.
For these reasons,

\begin{equation}
G_{sc}=+\infty.
\end{equation}
For any hopping $t$ and any $E_s$, $G_s$
is always less than $G_{sc}$.
Following discussions of {\bf Intermediate Hopping limit II} in
section VII, we therefore 
expect that when $G_c > G_{cc}$ or  $t > E_c G_{cc}$, ground 
states must be condensates of fractionalized chargons, 
either charge-e SSQC or charge-2e SSQC. These condensates 
are spin singlets, differing from
conventional pBECs discussed in high dimensional lattices (See Fig.9 and 
Fig.10).

The effective Hamiltonian for one-dimensional lattices can be written as
\begin{eqnarray}
&& {\cal H}^{1d}_{fqcb}=-\tilde{t}\sum_{\langle kl\rangle }({b}^{\dagger}_{k}{b}_{l}
\sigma^z_{kl} +h.c.)
\nonumber \\
&& + \sum_k \frac{{\rho}_k^2}{2C}-\mu\rho_k +\Gamma_b 
\sum_{\langle kl\rangle }\sigma^x_{kl}
\label{LEH}
\end{eqnarray}
which differs from ${\cal H}_{fqcb}$ in Eq.\ref{fqcb} by a term involving 
spatial plaquettes.
The Hilbert space of this Hamiltonian is subject to the same constraint
as Eq.\ref{constraintfb}

\begin{eqnarray*}
\hat{C}_k^{fb} \Psi=\Psi, \hat{C}_k^{fb}=\exp\big(i\pi [{b}^{\dagger}_k{b}_k
+\sum_{+}\frac{1-\sigma^x_{kl}}{2}]\big),
\end{eqnarray*}
and $\langle b^{\dagger}_kb_k\rangle =N_0$.

\begin{figure}
\begin{center}
\epsfbox{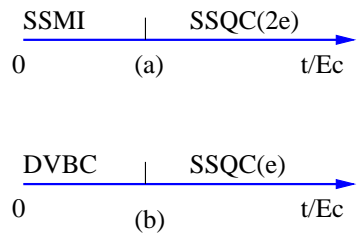}
\leavevmode
\end{center}
\caption{ States present in one-dimensional lattices with
an even or odd number of spin-one bosons per site
as the hopping $t$ is varied.
a) is for an even number of bosons per site;.
b) is for an odd number of bosons per site.
DVBC states in one-dimensional Mott states are
suggested by the quantum dimer model in section VIII;
in the limit $t \ll \sqrt{E_sE_c}$, these states are also
suggested by the bilinear-biquadratic $S=1$ spin chain model.
We believe these states appear in cold
atoms in an optical lattice when laser intensities are varied.
Note that SSQCs have only quasi-long range order in one-dimensional
limit and charge-e or -2e character might only be distinct close to
the
critical points.}
\end{figure}

Therefore, the physics of spin-one bosons 
with antiferromagnetic interactions in
one-dimensional lattices is effectively equivalent to the physics of
spinless bosons interacting with each other via Ising gauge fields. 
This is the key idea behind the notion of quantum condensates of
spin-one atoms; in this limit, spin-one atoms
behave as if they lost their identities as spinful particles.

Let us start with
an even number of particles per site. In this case,
the only fixed point of the Ising gauge field theory (unfrustrated) in 
one-dimensional lattices is a 
strong coupling fixed point regardless the value
of $\Gamma_b$.
The ground state for Ising fields should be unique with $\hat{T}_y=1$ 
and for any link $kl$,

\begin{equation}
\sigma^x_{kl}=1.
\end{equation}
And all excitations of Ising fields are fully gapped. 
The effective Hamiltonian can be reduced to ${\cal H}_{pb}$ in Eq.\ref{pb},
i.e.,

\begin{eqnarray}
&& {\cal H}^{1d}_{pb}=
-\tilde{t}_2\sum_{\langle kl\rangle }({b}^{\dagger}_{k}
{b}^{\dagger}_{k}{b}_{l}{b}_{l}+h.c.)
+\sum_k \frac{\rho_k^2}{2C}-\mu\rho_k,
\nonumber \\
&&( {\hat{C}^b_k})^{-1} {\cal H}_{pb} {\hat{C}^b_k}={\cal H}_{pb};
\end{eqnarray}
Therefore 
when $t_2 < E_c $, the ground state is a spin singlet Mott 
state, or
an SSMI;
furthermore, we believe that when $t_2 > E_{c}$, chargons are paired 
and "condense" forming a {\bf charge-2e} SSQC.
This leads 
to a phase diagram for one-dimensional optical lattices in Fig.9 a).

For an odd number of particles per site and when $E_c \gg E_s$,
one can carry out similar discussions. 
Following discussions in section VIII A and B, we find that the ground 
state for Ising 
fields 
has a two-fold degeneracy 
with $\hat{T}_y=\pm 1$, and for any link $kl$

\begin{equation}
\text{i)} \; \sigma^x_{kl}=(-1)^{k}, \; \mbox{or} 
\quad \text{ii)} \; \sigma^x_{kl}=(-1)^{k+1}
\end{equation}
corresponding to two states breaking the crystal translational symmetry.
And excitations are gapped.
In addition, because of the Berry's 
phase term (see discussions 
in section III), the Ising fields
are weakly interacting in this case.

Hopping of a chargon in this gauge field background is similar to hopping 
of an atom
in the projected spin singlet Hilbert space. A chargon hopps from
site $k$ to site $k+2$ via a virtual excitation involving an energy
of $E_s$. More specifically, in the representation introduced in section V
\begin{eqnarray*}
|...\rho_k,d_{k,k+1},\rho_{k+1},d_{k+1,k+2},\rho_{k+3},d_{k+3,k+4}...\rangle 
\end{eqnarray*}
one confirms that the hopping corresponds to the following process

\begin{eqnarray*}
&&|...2n+2,0,2n+1,1,2n+1,0...\rangle \rightarrow
\nonumber \\
&& |...2n+1,1,2n+2,1,2n+1,0..\rangle \rightarrow
\nonumber \\
&& |...2n+1,1,2n+1,0,2n+2,0..\rangle .
\end{eqnarray*}

Consequently, 
the effective Hamiltonian should be, instead of a paired hopping 
form 

\begin{eqnarray}
&& {\cal H}^{1d}_{fqb}=-\tilde{t}_2 \sum_{k}
(b^{\dagger}_kb_{k+2} + h.c.)+
\sum_k \frac{{\rho}_k^2}{2C}-\mu\rho_k.
\nonumber \\
\end{eqnarray}
At $G_c > G_{cc}$,
chargons condense and the ground state is a {\bf charge-e} SSQC,
similar to condensates of charged solitons proposed 
for non-integer numbers of atoms per site\cite{Zhou3}.
Subject to strong one-dimensional fluctuations,
SSQCs should be understood as states with 
quasi long-range phase order instead of condensates.

\begin{figure}
\begin{center}
\epsfbox{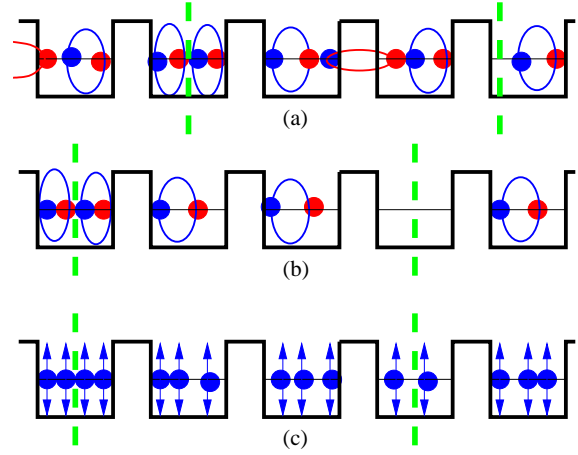}
\leavevmode
\end{center}
\caption{Condensation of particle-hole pairs suggested by
solutions to Eq.103. For an odd number of particles per site in a), a pair
of particles with $\pm$e charges are shown to condense;  in b)
for an
even number of particles per site, a pair of particles with $\pm$2e
charges condense. Note that these condensates are spin singlet states.
As a reference, in c) we also show condensation of particle-hole pairs in
a pBEC which is a stable phase only in high dimensional lattices.
Locations of charges are indicated by dashed lines.}
\end{figure}

The difference between condensates for odd and even numbers of atoms per 
site
can be attributed to the selection rule in low energy Hilbert 
spaces for each individual site defined in Eq.\ref{selec}.
In one-dimensional lattices,
the selection rule defined in section II leads to a two-fold degenerate 
DVBC for
an odd number of atoms per site and a non-degenerate SSMI for an even 
number of atoms per site in the strong coupling limit. Therefore, 
elementary charged excitations are
either charge-e ones or charge-2e ones depending
on the parity of numbers of atoms per site in one-dimensional lattices (see 
also Fig.4). Consequently,
condensates indicated in the fractionalized representation
involve condensation of chargons or chargon-pairs depending
on the parity of $N_0$.

\section{Possible Berry's phase effects on SSQCs in one dimensional 
lattices}

The action for Hamiltonian ${\cal
H}^{1d}_{\text{fqcb}}$ under the constraint defined after Eq.100
can be derived following a standard procedure; the result is

\begin{eqnarray}
&& S= - \sum_{\langle kl\rangle } J_{k,l} \sigma^z_{kl} \cos (\chi_{k}-\chi_{l})
\nonumber\\
&&
-k_\tau \sum^{\tau}_{\Box}\prod_{\Box}\sigma^z_{kl}+
i N_0\frac{\pi}{2}\sum_{k}(1-\sigma_{k k+\hat{\epsilon}}^z).
\nonumber \\
\end{eqnarray}
$J_{k,l}$ is given by

\begin{eqnarray}
&& J_{k, k+x}=(\epsilon t), J_{k, k+\tau}=(\epsilon E_c)^{-1} 
\end{eqnarray}
and $k_\tau$ is given in Eq.81.
This action should be used to study long wave length physics in 
one-dimensional 
lattices for an arbitrary hopping amplitude.

When $J_{k,l}$ is small, the phases ${\chi_k}$ are disordered;
these disordered phases represent Mott insulating states in one dimensional 
lattices 
which have been studied in details in section VIII.
From a point of view of the action in Eq.108, differences between
an even and odd number of atoms per site are only manifested in 
Berry's phase terms. 
The interesting "even-odd" effect in one-dimensional Mott states 
of spin-one bosons represents a well-known Berry's phase effect on spin 
liquids.

Results in the previous section appear to imply that 
properties of "Higgs" phases close to critical points
should also be solely determined by the even-odd parity of numbers of atoms per 
site.
Especially, for $N_0=2n$, the ordered phase represents a charge-2e
SSQC
and for $N_0=2n+1$, the ordered phase represents a charge-e SSQC.
Whether an one-dimensional
lattice with cold atoms perhaps is one of few systems where  
the Berry phase determines 
how the $U(1)$ symmetry is broken spontaneously remains to be understood.
In a separate paper, we are going to numerically study SSQCs 
and the speculated effect of Berry's phases on spontaneous symmetry
breaking.

\section{Conclusions}

In this article, we have illustrated the notion of fractionalized atoms 
and
demonstrated the possibility of having SSQCs of sodium 
atoms in 
one-dimensional optical lattices. 
Three major arguments have been developed to investigate
spin singlet Mott states and condensates. These are:

{\bf i}) In a projected spin singlet  
Hilbert space, spin-one 
bosons with antiferromagnetic interactions 
are equivalent to spinless bosons interacting with Ising 
gauge fields. 

{\bf ii}) Spin singlet Mott states are fully characterized by
even- and odd-class quantum dimer models.

{\bf iii}) Because of a selection rule in the Hilbert space, in 
one-dimensional 
lattices, superfluid phases of atoms can have either charge-e or 
charge-2e characters depending on the parity of numbers of bosons
per lattice site.

It is worth emphasizing that
{\bf charge-e} and {\bf charge-2e} condensates can be distinguished 
by studying frequencies in the ac Josephson effects.
Finally, the even-odd parity effect discussed here also appears to be 
attributed to
Berry's phase effects on spontaneous symmetry breaking.

Finally, we would like to mention two open questions which we believe
deserve further investigation. First, the nature of
states for non-integer numbers of particles per site is not clear to us;
possibilities of having spin singlet condensates in certain limits in this
case exist but need to be clarified.
There are also issues connected with phase separation, similar to
what happens in doped antiferromagnets\cite{Zaanen89,Schultz90}.
Second, it is appealing to verify results proposed in this article in 
the context 
of Luttinger-
liquid theories; at the moment we are not aware of 
such attempts and
do not know if results discussed in this paper can be rederived
in a Luttinger-liquid theory based approach. 
Although it is not clear to us whether the phenomena
discussed in this article are related to the Luttinger-liquid physics,
attempts along this line of thought might shed light on the further 
understanding of one-dimensional
phases of spin-one bosons with antiferromagnetic interactions.

\section{acknowledgement}

We would like to thank the Lorentz center, Leiden
for its hospitality during the 2002 workshop on "quantum spin collective
phenomena in solid state physics"; 
FZ also wants to thank the Aspen center for physics
during the 2003 workshop on "quantum gases"
where this work was finished.
Finally, one of us (FZ) wishes to thank P. W. Anderson
for his interest on this subject,
F. D. M. Haldane and
P. B. Wiegmann for many stimulating discussions on SSQCs. 
This work is supported by the Foundation FOM, the Netherlands
under contracts 00CCSPP10, 02SIC25 and NWO-MK "projectruimte" 00PR1929;
MS was also partially supported by a grant from Utrecht university.

\appendix

\section{charge-2e excitations in DVBCs in square lattices}

For a DVBC in square lattices,
the wave function can also be represented by Eq.43;
the product is however carried over all even (odd) links in each row or 
in each column. The ground state thus has a fourfold 
degeneracy\cite{Zhou3}. 

The excitations in 2d DVBC states have some fascinating 
properties. 
It is convenient to employ the following modified creation and annihilation
operators of valence bonds,

\begin{eqnarray}
&& \tilde{\phi^{\dagger}}_\eta=
\phi^{\dagger}_\eta\bar{\delta}(\eta), \tilde{\phi}_\eta=\phi_\eta\delta(\eta).
\end{eqnarray}
where $\phi^{\dagger}_\eta$, $\phi_\eta$ are the creation and annihilation 
operators
of spin singlet states of two coupled condensates defined in 
Eq.\ref{creation}
at a link $\eta$
and

\begin{equation}
\delta(\eta)= \left\{ 
\begin{array}{cc} 
1, & \mbox{if $\phi^{\dagger}_\eta\phi_\eta\Psi=\Psi$;} \\
0, & \mbox{otherwise}
\end{array} 
\right.
\end{equation}
Finally,
$\bar{\delta}(\eta)=1-\delta(\eta)$.
The reduced Hamiltonian can be written as

\begin{equation}
{\cal H}_{rd}=\sum_{\eta_1,\eta_2}[ V\tilde{\phi}^{\dagger}_{\eta_1}
\tilde{\phi}_{\eta_1}
\tilde{\phi}^{+}_{\eta_2}\tilde{\phi}_{\eta_2}
-T \left(\tilde{\phi}^{\dagger}_{\eta_1}
\tilde{\phi}^{\dagger}_{\eta_2}
\tilde{\phi}_{\bar{\eta}_1}\tilde{\phi}_{\bar{\eta}_2}+h.c.\right)]
\end{equation}
$\eta_1, \eta_2$ are a pair of two parallel {\em adjacent} links,
perpendicular to another pair $\bar{\eta}_1$, $\bar{\eta}_2$ in an 
elementary plaquette. 

To facilitate discussions, we introduce 
a local gauge transformation $\hat{E}_k$ at site $k=(m,n)$;

\begin{equation}
\hat{E}_{k}=\exp\big(i\pi[\sum_{+}\tilde{\phi}^{\dagger}_\eta\tilde{\phi}_\eta-1 
]\big)
\end{equation}
where the sum is again over all neighboring sites of site $k$.
It is easy to confirm that all the resonating valence bond 
configurations
and the Hamiltonian are invariant under this local gauge transformation

\begin{equation}
\hat{E}_k \Psi=\Psi, \hat{E}_k^{-1} {\cal H}_{rd} \hat{E}_k={\cal 
H}_{rd}.\label{inv}
\end{equation}

To facilitate discussions, we consider a limit 
where two parallel valence bonds attract each other strongly
and $-V$ is much greater than the exchange interaction $T$. 
(The following 
discussions are also expected to be 
valid as far as $-V < T$ because a bare $-V$ is always renormalized 
to infinity in this limit.)
By making an expansion over $T/|V|$, one anticipates the following
ground state wave function,

\begin{eqnarray}
&& |G^o_1\rangle =|CS_1\rangle 
+ \sum_M \big(\frac{{\cal V}}{V}\big)^M |CS_1\rangle 
\nonumber \\
&& {\cal 
V}=-T\sum_{\eta_1,\eta_2}\tilde{\phi}^{\dagger}_{\eta_1}
\tilde{\phi}^{\dagger}_{\eta_2}
\tilde{\phi}_{\bar{\eta}_1}\tilde{\phi}_{\bar{\eta}_2}+h.c.
\label{per}
\end{eqnarray}
where $|CS_1\rangle $ is one of the four fold 
degenerate column states(see discussions at the beginning of this section).

Let us start with solitonic spin-one excitations.
Such a spin-one excitation is created by 
breaking one valence bond
and sending one of the unpaired condensates 
to infinity while keeping the other one fixed.
The many-body wave function of a solitonic excitation localized at site $k$ 
would 
thus be
created by the operator

\begin{eqnarray}
&& \Psi^{S+}_{\cal M}(k)={\cal P}^1_{G.G.}{\bf h}^{\dagger}_{k,\alpha} 
\prod_{\eta_e \in 
{\cal M}}
\tilde{\phi}^{\dagger}_{\eta_e}
\prod_{\eta_o \in {\cal M}}
\tilde{\phi}_{\eta_o} 
\end{eqnarray}
where ${\bf h}^{\dagger}_{k}$ is a creation operator defined at site $k$.
${\cal M}$ is a path starting at site $k$ and terminated at infinity
and $\eta_{o,e}$ are the odd and even links along the path 
(counted 
from site $k$).
In one-dimensional lattices, 
this precisely creates a spin-one kink as shown in 
\cite{Zhou3}.

In a leading order of $T/|V|$, we estimate the energy of the spin-one 
excitation 

\begin{equation}
E=\frac{\langle G^o_1|\Psi_{\cal M} {\cal H}_{rd} \Psi^{\dagger}_{\cal M}|G^o_1\rangle }
{\langle G^o_1|\Psi_{\cal M}\Psi^{\dagger}_{\cal M}|G^o_1\rangle } = V L_{\cal M}
\end{equation}
where $L_{\cal M}$ 
scales as the size of the system and the excitation is infinitely
massive in a thermodynamical limit.

\begin{figure}
\begin{center}
\epsfbox{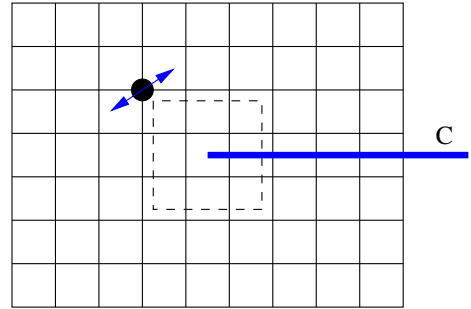}
\leavevmode
\end{center}
\caption{Moving an $S=1, Q=0$
excitation (a dot with a double headed arrow) around a dual
particle (defined along path ${\cal C}$, see Eq. A11)
results in a minus sign in the many-body
wave
function;
a dual particle and an $S=1, Q=0$ excitation view each other as half
vortices.}
\end{figure}

The absence of spin-one solitonic excitations in the low energy 
excitation 
spectra 
doesn't depend on the particular form of wave functions we used.
It is a topological property of the ground state which determines the 
interactions between solitonic excitations.
Here we are going to
provide some understanding based on the duality between spin 
excitations
which are "electric" and
topological excitations with dual "magnetic" charges.

To identify this gauge transformation $\hat{E}_k$ as the one of Ising 
gauge fields, one notices typically

\begin{eqnarray}
&& \hat{E}^{-1}_{k} 
\tilde{\Sigma}^z_{\eta}\hat{E}_{k} =
(-1)^{\pi(\eta,k)} \tilde{\Sigma}_{\eta}^z, 
\tilde{\Sigma}^z_{\eta}=\tilde{\phi}^{\dagger}_\eta-\tilde{\phi}_\eta.
\end{eqnarray}
Here $\pi(\eta,k)$ is unity if link $\eta$ is connected with site $k$;
otherwise it is zero.
As $\tilde{\Sigma}^z$ satisfies the algebras of Pauli matrices
and should be identified as $\sigma^z$ of the Ising gauge fields,
the local gauge transformation effectively yields $\sigma^z \rightarrow 
-\sigma^z$
and generates a desired Ising gauge transformation.

An operator carries an "electric" charge if it transforms nontrivially 
under the local gauge transformation. In our case, we can show that

\begin{equation}
\hat{E}^{-1}_{k}
\Psi^{S+}_{\cal M}(k') \hat{E}_k=
(-1)^{\delta(k-k')} \Psi^{S+}_{\cal M}(k');
\end{equation}
and the creation operator of a spin one solitonic excitation carries a 
charge
defined with respect to the local gauge transformation.
The corresponding particle thus is topological in nature and violates
the invariance defined in Eq.\ref{inv} only at the site $k'$ where the 
particle 
is created.

Furthermore, we can introduce
a creation operator of a dual magnetic charge;

\begin{eqnarray}
&& \Psi^{e+}_{\cal C}(x,y)=\prod_{\eta \in {\cal C}}
\big(2 \tilde{\phi}^{\dagger}_\eta
\tilde{\phi}_\eta - 1 \big) 
\nonumber \\
\end{eqnarray}
where the path ${\cal C}$ begins at the center of a plaquette $(x,y)$ and
ends at infinity by piercing all links intersecting with it.
In the leading order of $T/|V|$, one also finds that

\begin{equation}
\Psi^{e+}_{\cal C} |G_1^o\rangle \approx (-1)^{N_{\cal M}} |G_1^o\rangle 
\end{equation}
where ${N_{\cal M}}$ is the number of unoccupied links 
along the path ${\cal M}$.
This implies that the ground state should be a condensate of dual 
particles.

Now one can move 
a spin-one solitonic spin 
excitation $\Psi^{\dagger}_{\cal M}$ around
a dual magnetic charge ${\Psi^{\dagger}_{\cal C}}$.
In the Hilbert space spanned by valence bond configurations, the following 
commutation relation holds:

\begin{eqnarray}
&& \{ \big(2 \tilde{\phi}^{\dagger}_\eta
\tilde{\phi}_\eta - 1 \big),
\tilde{\phi}^{\dagger}_\eta\}=
\{ \big(2 \tilde{\phi}^{\dagger}_\eta
\tilde{\phi}_\eta - 1 \big),
\tilde{\phi}_\eta\}=0.
\label{anticom}
\end{eqnarray}
Following Eq. \ref{anticom}, one verifies
that the many-body wave function changes sign after this operation
(shown in FIG.11).
This implies remarkable long range correlations in the system.
Since the "magnetic" and "electric" charges
in the theory see each other as half-flux-quanta , it is imaginable that
as one of the charges condenses, the other one gets confined
following a standard picture in topological field theories.
In the current context, the dual magnetic charges condense and 
spin-one solitonic spin excitations are confined because of random 
phases developed when moving in a dual magnetic charge condensate.

We are now ready to address the quantum numbers of {\em elementary} spin 
excitations. Because of confining forces between $S=1$ solitonic 
excitations, only bound states of soliton-anti soliton pairs can exist in 
the excitation spectrum as elementary excitations.
So there will be two branches spin excitations with $S=1$ and $S=2$ 
respectively.
Taking into account Eq.\ref{spectra}, we should further expect the bound 
states of $S=2$ have lower energies than $S=1$.
In fact, microscopically these bound states are likely to be the excited 
states of $S=2$ and 
$S=1$ of two coupled neighboring condensates.

A {\bf charge-e} soliton can be created by adding an atom to a site while
keeping the rest of sites paired; this excitation is a spin singlet as
the ground state but with an extra atom or {\bf charge-e}.
For the same reason discussed before for a spin-one solitonic excitation, 
a charged soliton is infinitely 
massive because of the long range rearrangement of valence bonds
and intends to pair with another charged soliton. 
Naturally, at low energies, one can only have {\bf charge-2e} spin singlet 
excitations in this case.

\section{The derivation of the fractionalized representation}

The equivalence between Eq. 6 and Eq.\ref{Hami} can be demonstrated by 
examining the partition function,

\begin{equation}
{\cal Z}=Tr[\exp(-\beta {\cal H}_{FR}) {\cal P}];
\end{equation}
${\cal P}$ is a projection operator defined in Eq.(B6)

To facilitate the calculation, we introduce the following 
special "coherent"
state representation

\begin{eqnarray}
&& |{\bf n}\rangle =
\frac{1}{\sqrt{2\delta N}} 
\sum_{N_0-\delta N}^{N_0+\delta N}\frac{({\bf 
n}\cdot {\bf a}^{\dagger})^n}{\sqrt{2(n-1)!}} |0\rangle ,
\nonumber \\
&& |z\rangle =
\frac{1}{\sqrt{2\delta N}}
\sum_{N_0-\delta N}^{N_0+\delta N}
\frac{(z b^{\dagger})^n}{\sqrt{2(n-1)!}}|0\rangle 
\end{eqnarray}
assuming that $N_{0} \gg \delta N$ but both are much larger than unity 
and $z=\exp(i\phi)$. One can easily show that at large $N$ limit,

\begin{eqnarray}
\langle {\bf n}|{\bf n}'\rangle \approx \delta({\bf n}-{\bf n'}),
\langle z|z'\rangle \approx \delta(z-z');
\end{eqnarray}

and

\begin{eqnarray}
&& \rho |z\rangle =\frac{\partial}{\partial \ln z}|z\rangle  \nonumber \\
&& {\bf S}|{\bf n}\rangle =i{\bf n}\times \frac{\partial}{\partial
{\bf n}}|{\bf n}\rangle .
\label{coherent}
\end{eqnarray}
For the study of the action, we slice the ($d$+1) dimensional Euclidean
space into $M$ slices
along the temporal direction and rewrite the partition function as

\begin{eqnarray}
&& {\cal Z}=\prod_\tau
\langle \{ {\bf n}_{k,\tau} \},
\{ z_{k,\tau}\}, \{\sigma^z_{k,\tau}\}|
\exp(-\epsilon {\cal H}_{FR}) {\cal P} 
\nonumber \\
&& |\{ {\bf n}_{k,\tau +\epsilon} \},
\{ z_{k,\tau +\epsilon}\}, \{\sigma^z_{k,\tau +\epsilon}\}\rangle .
\nonumber \\
\label{action}
\end{eqnarray}
Here $\tau$ and $\tau+\epsilon$ label two adjacent slices and
$\epsilon=\beta/M$.
The projection operator is to impose a constraint due to the symmetry of
many-boson wave functions,

\begin{equation}
{\cal P}=\sum_{\xi_k=\pm 1}\exp[i\frac{1+\xi_k}{2}\pi(n_k+S_k)]
\end{equation}
 so that only states with an even $n_k+S_k$ contribute to the functional
integral.

The matrix element in Eq.\ref{action} can be conveniently evaluated
by inserting a complete set of $n_k, S_k$, the eigen states of the
number and spin operators.

\begin{eqnarray}
&& \sum_{\{n_k\},\{S_k\}}\langle \{ {\bf n}_{k,\tau} \},
\{ z_{k,\tau}\}, \{\sigma^z_{k,\tau}\}|
\exp(-\epsilon {\cal H}_{FR}) {\cal P} \otimes
\nonumber \\
&& |\{n_k\}, \{S_k \}\rangle 
\langle \{n_k \}, \{S_k\}|\{ {\bf n}_{k,\tau +\epsilon} \},
\{ z_{k,\tau +\epsilon}\}, \{\sigma^z_{k,\tau +\epsilon}\}\rangle .
\end{eqnarray}
Here $z_k=\exp(i\chi_k)$.

Consider a simplified situation of a planar ${\bf n}=
(\cos\phi,\sin\phi,0)$ ( a generalization appears to be possible).
As suggested in Eq.\ref{coherent}, ${\bf S}$ and $\rho$ are conjugate 
variables of ${\bf n}$ and $\phi$.
So we can express the
the eigen states $|n_k\rangle $, $|l_k\rangle $ in the following simple forms

\begin{eqnarray}
&&\langle z_k|n_k\rangle =\frac{1}{\sqrt{2\pi}}\exp(i\chi_k n_k),
\nonumber \\
&& \langle {\bf n}_k|S_k\rangle =\frac{1}{\sqrt{2\pi}}\exp(i\phi_k S_k).
\end{eqnarray}

The rest of the derivation is identical to that in \cite{Demler01}.
Redefining $\xi_k$ as $\sigma^z$ along a link located at site k between
slice $\tau$ and $\tau+\epsilon$ and summing up all contributions at
different slices, we obtain the result in the Euclidean space 
as

\begin{eqnarray}
&& S= - \sum_{rr'} J^c_{rr'} \sigma^z_{rr'} cos \chi_{rr'}
   - \sum_{rr'} J^{2c}_{rr'}  \cos (2 \chi_{rr'})
\nonumber\\
&&   - \sum_{rr'} J^s_{rr'} \sigma^z_{rr'} {\bf n}_r {\bf 
n}_{r'}
  - \sum_{rr'} J^{2s}_{rr'}  Q^{ab}_r Q^{ab}_{r'}
\nonumber \\
&& +iN_0 \sum_{r}\frac{1-\sigma_{r r_\tau}}{2}\pi 
\label{Sf}
\end{eqnarray}
Here summation goes over sites $r=(k,\tau)$ in the space-time lattice,
$\chi_{rr'}=\chi_r-\chi_{r'}$, and

\begin{eqnarray*}
Q^{ab}_r = {\bf n}^a_r
{\bf n}^b_r - \delta^{ab}/3
\end{eqnarray*}
is a nematic order
parameter, and $\sigma^z_{rr'}= \pm 1$ is an Ising field
that lives on the links and
$r_\tau=r+\hat{\epsilon}$.

The coupling constants
are 

\begin{eqnarray*}
&& J^c_{r,r \pm \hat{\tau}}=\frac{1}{E_c\epsilon},
J^s_{r,r \pm \hat{\tau}}=\frac{1}{E_s\epsilon},
J^{c,s}_{r,r \pm \{\hat{x},\hat{y},\hat{z}\}}=\epsilon t N_0,\nonumber \\
&& J^{2c}_{r,r \pm \hat{\tau}}=J^{2s}_{r,r \pm \hat{\tau}}=0,
J^{2c,2s}_{r,r \pm \{\hat{x},\hat{y},\hat{z}\}}=-\epsilon t N_0 /4\nonumber \\.
\end{eqnarray*}
This is precisely the action of Eq. 1 derived in an early
work\cite{Demler01}(the Berry's phase term was omitted there). Therefore 
we established an
anticipated equivalence between Eq. 1 and Eq.\ref{Hami}.
Finally, we would like to mention that
Ising-gauge-theory based approaches have also been employed to study 
electron fractionalization 
in previous works 
\cite{Read91,Jalabert91,Senthil99,Senthil00,Moessner01,Wen91,Balents98,Sachdev00}.

\section{NMIs in high dimensional lattices: Revisit I}

As chargons are fully gapped, states in this limit are 
incompressible.
Integrating out chargons' or $b$-particles' degree of freedom in 
Eq.\ref{Hami2} 
we end up with an effective Hamiltonian

\begin{eqnarray}
&& {\cal H}_{fqca}=
-\tilde{t}\sum_{\langle kl\rangle }({\bf a}^{\dagger}_{k\alpha}{\bf
a}_{l\alpha}\sigma^z_{kl} +h.c.)+
\sum_k \frac{{\bf S}^2_k}{2I}
\nonumber \\
&& +\Gamma_a \sum_{\langle kl\rangle }\sigma^x_{kl}-K_{sa}
\sum_{\Box}\prod_{\Box}\sigma^z_{kl}.
\end{eqnarray}
Calculations indicate that
$\Gamma_a, K_{sa}$ should be functions of $t, E_c$.
When $t$ is much smaller than $E_c$, or $G_c=tE_c^{-1}$ is small,
the ratio $K_{sa}\Gamma_a^{-1}$ is much less than unity,
or
\begin{equation}
\frac{K_{sa}}{\Gamma_a} \ll 1 (\Gamma_a\sim E_c). 
\label{ksaovergamma}
\end{equation}
However, we speculate that close to the critical value $G_{cc}$, the ratio in 
Eq. \ref{ksaovergamma}
should be divergent (one should also expect terms involving large loops).
Finally, the integration also leads to a
new constraint on the Hilbert space,

\begin{eqnarray}
\hat{C}_k^{fa} \Psi=\Psi, \hat{C}_k^{fa}=\exp\big(i\pi [{\bf a}^{\dagger}_k{\bf 
a}_k
+\sum_{+}\frac{1-\sigma^x_{kl}}{2}]\big).
\end{eqnarray}
The sum over "+" is again carried over all links connected to site $k$.
And the Hamiltonian is locally invariant under the action of
$\hat{C}^{fa}_k$

\begin{equation}
\big(\hat{C}_k^{fa}\big)^{-1} {\cal H}_{fqca} \hat{C}_k^{fa}=
{\cal H}_{fqca}.
\end{equation}

The condensation can be easily visualized
when $t$ is close to $E_{c}G_{cc}$, or   
in the large $K_{sa}$ limit.
The gauge fields are weakly interacting and fluctuations are small. For
instance, one can estimate the probability of finding a loop of plaquettes
with $\prod_{\Box}\sigma^z_{kl}=-1$ in (2+1) Euclidean space
is exponentially small.

It is therefore tempting to work out a self-consistent solution
in a "mean field" approximation where $\prod_{\Box}\sigma^z_{kl}=1$ at
each plaquette. Furthermore, one chooses a unity gauge

\begin{equation}
\sigma^z_{kl}=1, \mbox{for any link}.
\end{equation}

One then obtains a FQCa, or an ${\bf a}$-type FQC, as 
the
ground state
in this case. In FQCa,

\begin{equation}
{\cal G}^b(k,k+\infty)=0,
{\cal G}^{\bf a}_{\alpha\beta}(k,k+\infty)\sim {\bf
n}_\alpha{\bf n}_\beta
\end{equation}
as spin-one spinons condense in a state created by an operator
${\bf a}_{Q=0}^{\dagger} \cdot {\bf n}$.

It is also possible to show the existence of the fractionalized phase
beyond this simple mean field theory. In a dilute gas approximation
suggested in \cite{Fradkin79}, one finds an up-bound of the correction to
${\cal G}^{\bf a}_{\alpha\beta}(k, k+\infty)$ calculated in a mean field
approximation which is exponentially small as $K_{sa}$ or $G_s$ is large.
One therefore expects that the conclusions arrived above are valid beyond 
the
mean field approximation.

The ground state of FQCa does not exhibit phase coherence or
Josephson effects.
However, the rotational symmetry is broken.
There are two branches gapless spin wave excitations (spin one);
the zeroth sound
involving the compression or expansion of particle-densities is fully
gapped.

A more tricky situation is
when $t$ is much less than $E_c$
or correspondingly when $\Gamma_a$ is
large and gauge   
fluctuations are strong in the absence of matter fields.
To argue the existence of condensation in this case, one 
recalls that there are no phase transitions when $K_a$ is varied but
$G_s$ is kept at infinity.
Physically,
presence of matters strongly renormalizes gauge field
dynamics.
A condensate represents a self-consistent solution to the
problem provided $t$ is much larger than $E_s$;
this conventional point of view is supported by 
results in \cite{Toner}.
However, since pure gauge fields at large $\Gamma_a$ limit are confining,
only pairs of ${\bf a}^{\dagger}{\bf a}^{\dagger}$ are free excitations at small $t$
limit; as $t$ is increased, condensation appears to only involve   
these pairs\cite{pFQCa,Sugar}. To illustrate this point, we
present some detailed discussions on
pFQCa.

As stated at the beginning of this section,
there are even numbers of atoms
at each site. Consequently,
each site is occupied by an even number of chargons;
excitations involving a change of numbers of chargons at each site
are gapped by an energy of order of $E_c$.
At energies much lower than $E_c$,
the constraint in Eq.\ref{constrain} then indicates that the low energy
Hilbert space should be spanned by states
satisfying

\begin{equation} 
\hat{\cal C}_k^a \Psi=\Psi,
\hat{\cal C}_k^a=\exp\big(i\pi{\bf a}^{\dagger}_k\cdot{\bf a}_k\big).
\end{equation}

In this subspace defined by two-particle operators $({\bf a}_k^{\dagger})^2$, the
relevant Hamiltonian is due to pair hopping
following Eqs.\ref{Hami},\ref{Hami2},

\begin{eqnarray}
&& {\cal H}_{pa}=
-\tilde{t}_1\sum_{\langle kl\rangle }({\bf a}^{\dagger}_{k\alpha}
{\bf a}^{\dagger}_{k\beta}{\bf a}_{l\alpha}{\bf a}_{l\beta}
+ h.c.)
+\sum\frac{{\bf S}_k^2}{2I}
\nonumber \\
&& \big( \hat{C}^a_k \big)^{-1} {\cal H}_{pa} {\hat{C}^a_k}={\cal
H}_{pa}.
\label{Hpa}
\end{eqnarray}
and

\begin{eqnarray} 
\tilde{t}_1 \approx \frac{{t}^2}{\Gamma_a} (\Gamma_a\sim E_c).
\end{eqnarray}
By examining the Hamiltonian in this reduced space,
assuming $t_1=\tilde{t}_1N_a^2 \gg E_s $,
one concludes that the ground state is a
condensate
of paired spinons, or pFQCa.
In a lattice with $V_T$ sites, the wave function of a
pFQCa
can be
written as (up to a normalization factor)

\begin{eqnarray}
&& |g_1\rangle =\prod_k \frac{b_k^{+{2n}}}{\sqrt{2n!}} \otimes
{\cal P}_{N_a\times V_T}\exp\big(
\Phi^{\dagger}_a(Q_0=0)\big) |0\rangle .
\nonumber \\
&& \Phi^{\dagger}_a(Q_0)= {\bf Q}_{\alpha\beta}
\sum_q {\bf a}^{\dagger}_{q,\alpha}{\bf a}^{\dagger}_{-q+Q_0,\beta}
\label{pFQCa}
\end{eqnarray}
which explicitly satisfies the above constraint.
Here $\Psi^{\dagger}_a(Q_0)$ is a creation operator for a pair of spinons
in $(Q_0-q, q)$ channel.
The sum over $q$ is carried over the first Brillouin zone.
${\cal P}_{N_a\times V_T}$ is to project out 
$N_a \times V_T$-particle states.

In a pFQCa, ${\cal G}^{\bf a}_{\alpha\beta}(k,l)$ and ${\cal G}^b(k,l)$
vanish as $k-l$ approaches
infinity.
However, the
rotational symmetry is still broken and
there exists the following long range order in the
generalized BCS wave function

\begin{equation}
{\Delta}^{\bf a}_{\alpha\beta}(k,k)
=3 {\bf Q}_{\alpha_\beta}
\end{equation}
which is invariant under the local gauge transformation defined in
Eq.\ref{tran}.

The difference between pFQCa and FQCa is subtle.
Both of them break rotational symmetry, support gapless modes and
have similar local dynamics. On the other hand, topological excitations in
two states are distinct.

Let us also mention briefly
that when $E_s \gg t,t_1$, only localized singlet pairs
of spinons are allowed in the system.
These states can be identified
as SSMIs discussed in section II.

\section{DVBC'S in a fractionalized representation}

As shown in some details, the ground state for an odd $N_0$ when 
$G_{s,t} 
\ll 1$ is
a projected superposition of paired condensates in $(q, -q)$ and
$(q, \pi-q)$ channels and has a two-fold degeneracy. 
The corresponding wave functions 
of the two-fold degenerate ground states are topologically identical to

\begin{eqnarray}
&& |g_3\rangle =\prod_k (b^{\dagger}_k)^{2n+1} \otimes \frac{({\bf a}^{\dagger}_{k\alpha}{\bf 
a}^{\dagger}_{k\alpha})^{n}}{\sqrt{(2n+1)!}} \nonumber \\
&& {\cal P}^{1}_{G.G.}
\exp \lbrack \sum_{\gamma=0,1} g_{\gamma}
\Phi_{as}^{\dagger}\big(\gamma \pi\big) \rbrack |0\rangle ;
\nonumber \\
&& \Phi^{\dagger}_{as}({Q}_0)=
\sum_q\frac{h({q})}{2\sqrt{3}} {\bf a}^{\dagger}_{{Q_0+q},\beta}
{\bf a}^{\dagger}_{-q,\beta}
\label{dVBC}
\end{eqnarray}
where $h({q})$ is chosen to be $\exp(i{q})$.
Two fold degenerate states correspond to $g_0=g_1=1$ or $g_0=-g_1=1$. 
It is straightforward to establish an equivalence between 
this projected pFQCa state and dimerized valence bond crystals
studied previously\cite{Zhou3}.

\section{The Derivation of Eq.80}

Let us consider a limit where $\chi$ and ${\bf n}$ have short
range correlations and 

\begin{equation}
\langle cos\chi_r \cos\chi_l\rangle =C_\chi,
\langle {\bf n}_r {\bf 
n}_l\rangle=C_{\bf n}.
\end{equation}
only if $k, l$ are two neighboring sites.

Using the standard high temperature expansion technique, one integrates
out ${\bf n}, \chi$ in the action given in Eq.B9. The results
in the leading order are

\begin{eqnarray}
&& {\cal Z}(\{\sigma_i^z\})\sim 
\prod_{i}\sum_{\sigma_i^z=\pm 1}\{1
\nonumber \\
&&-(\epsilon t)^4 \sum^s_{kl,lm,mn,nk} 
\sigma^z_{kl}\sigma^z_{lm}
\sigma^z_{mn}\sigma^z_{nk}\\
&& -[(\frac{t}{E_c})^2 +(\frac{t}{E_s})^2]
\sum^\tau_{kl,lm,mn,nk} \sigma^z_{kl}\sigma^z_{lm}
\sigma^z_{mn}\sigma^z_{nk} \}\\
&&\times \exp (iN_0 \sum_{r}\frac{1-\sigma^z_{r r_\tau}}{2}\pi).
\end{eqnarray}
The sum in Eq.E2 is over all elementary spatial plaquettes occupied by 
links
$kl$,$lm$,$mn$,$nk$;
the sum in Eq.E3 is over elementary plaquettes involved both spatial 
links $kl$, $mn$ and temporal links $lm$, $nk$; without losing generality,
we have set $C_{\chi}$ and $C_{\bf n}$ to be unity.
The structure of Eq.E2 is identical to Eq.80; further more,
comparing Eqs.E2,E3,E4 with Eq.80, one obtains results in Eq.81.

\section{Gauge invariant correlation functions}

It is tempting to classify SSQCs or more general FQCs in terms of local 
order
parameters similar to those in Eq.1. To characterize FQCs, in addition 
one should introduce

\begin{eqnarray}
&& {\cal O}^0=\langle \psi_{k\alpha}\psi_{k\alpha}\rangle . \nonumber \\
\end{eqnarray}
In terms of order parameters ${\cal O}^{0,1,2}$,

\begin{eqnarray}
&&\mbox{   pBEC:    } {\cal O}^1_\alpha \neq 0, {\cal 
O}^2_{\alpha\beta}\neq 0, {\cal O}^0\neq 0;\nonumber \\
&&\mbox{  SSQC(2e):   } {\cal O}^1_\alpha =0, {\cal O}^0 \neq 0,
{\cal O}^2_{\alpha\beta}=0;
\nonumber \\
&&\mbox{ pFQCa:   } 
{\cal O}^1_\alpha =0, {\cal O}^2_{\alpha\beta}\neq 0, {\cal O}^0=0.
\end{eqnarray}
For FQCa and SSQC(e) states, 
the order parameters 
are nonlocal in terms of $\psi^{\dagger}$ and will be introduced in a 
subsequent 
paper on fractionalized states for non-integer numbers of bosons per 
site.

We would like to make two more 
remarks on FQCs. First,
Elitzur theorem claims that in the presence of local gauge invariance,
operators such as $\sigma^{\pm}_k$,
${\bf a}^{\dagger}_k$, $b_k^{\dagger}$ which transform nontrivially under the 
local gauge transformation can not develop
nonzero expectation values\cite{Elitzur75}.
The existence of FQCs 
which so far is explicitly demonstrated in a fixed gauge 
at first sight appears to be at odds with the Elitzur theorem.
The apparent paradox can be formally resolved if one carries out a 
similar 
calculation in terms of gauge invariant correlators and confirms 
long range order.

Indeed, for discussions on quantum condensates,
one can also study the following gauge invariant two-point correlation 
functions

\begin{eqnarray}
&& G^{\bf a}_{\alpha\beta}(k,l)=
\langle {\bf a}^{\dagger}_{k\alpha} 
\prod_{C_{kl}}\sigma_t^z 
{\bf 
a}_{l\beta} \rangle  
\langle \prod_{C^4_{kl}}\sigma_t^z\rangle ^{-1/4}  \nonumber \\
&& {G}^b(k,l)=\langle {b}^{\dagger}_k 
\prod_{C_{kl}}\sigma_t^z {b}_l \rangle 
\langle \prod_{C^4_{kl}}\sigma_t^z\rangle ^{-1/4}  \nonumber \\
\end{eqnarray}
where the products are carried over links along path ${C_{kl}}$
and $C^4_{kl}$. $C_{kl}$ is a path connecting $k$ and $l$ site and
$C^4_{kl}$ is a closed path of four times as long.
These correlation functions characterize the condensates 
when the gauge fields are weakly interacting\cite{gauge}.

Gauge invariant correlators suggest that
it should not be bare ${\bf a}$- or $b$-particles but some nonlocal 
"particles"
which condense. One such candidate is an ${\bf a}$ or $b$-particle
dressed in a string of gauge fields

\begin{equation}
{\bf A}^{\dagger}_{k\alpha}(C_k)={\bf a}^{\dagger}_{k\alpha} \prod_{C_k}\sigma^z_\eta,
{B}^{\dagger}_{k}(C_k)={b}^{\dagger}_{k} \prod_{C_k}\sigma^z_\eta
\end{equation}
where $C_k$ is a path starting at site $k$ and terminated at 
infinity.
One can easily confirm the following local gauge invariance of these 
operators

\begin{eqnarray}
&&\big(\hat{C}^{fa}_{k}\big)^{-1}{\bf A}^{\dagger}_{k\alpha}
\hat{C}^{fa}_{k}
={\bf A}^{\dagger}_{k\alpha},\nonumber \\
&&
\big(\hat{C}^{fb}_{k}\big)^{-1}{B}^{\dagger}_{k}
\hat{C}^{fb}_{k}
={B}^{\dagger}_{k}.\nonumber \\
\end{eqnarray}
These particles can condense without violating the Elitzur theorem.
On the other hand they carry exactly the same charge and spin
as bare particles of ${\bf a}^{\dagger}$ and $b^{\dagger}$. 
We speculate that
condensation of ${\bf a}$ or $b$-particles in a fixed gauge 
might also imply condensation of particles created by ${\bf A}^{\dagger}$ 
and $B^{\dagger}$.

The second remark concerns the relation between charge-e SSQCs and 
previously 
discussed spin singlet paired condensates
or charge-2e SSQCs\cite{Demler01}.
Both condensates are rotationally
invariant spin singlets and both are phase coherent. However,
from the point of view of correlations it is obvious that 
charge-e SSQC doesn't involve pairing of particles and differs from paired 
condensates. 
An explicit construction of wave functions of charge-e SSQC
is difficult but available.
At the time of writing, we believe that FQCs represent condensation
of some topological solitons which in short are called as
${\bf a}$-, $b$- particles. Particularly, charge-e SSQCs should 
exist in 
lattices with both non-integer and integer 
numbers of atoms per site.

\end{multicols}


\begin{thebibliography}{99}
\bibitem{pFQCa}
However, it is not clear how phase transitions between
pFQCa and FQCa states take place.
\end{thebibliography}
\end{document}